\documentclass[12pt]{article}

\usepackage{mathptmx}
\usepackage{setspace}
\usepackage[margin=1.0in]{geometry}
\usepackage{authblk}
\usepackage{abstract}
\usepackage{amsmath}
\usepackage[usenames,dvipsnames,svgnames,table]{xcolor}

\usepackage{subcaption}

\usepackage{graphicx}
\usepackage{booktabs}
\usepackage{cleveref}

\usepackage{verbatim}
\usepackage{anyfontsize}
\usepackage{nameref}
\usepackage{nicefrac}
\usepackage{float}
\usepackage{bm}
\usepackage{multirow}
\usepackage[ruled,vlined,linesnumbered]{algorithm2e}
\usepackage{tikz}
\usepackage[group-separator={,}]{siunitx}
\usepackage{xspace}

\usepackage{hyperref}

\newcommand{\forexample}{e.g.\@\xspace}
\newcommand{\thatis}{i.e.\@\xspace}

\newcommand{\andothers}{et al.\@\xspace}
\newcommand{\totCount}[1]{\ensuremath{T(#1)}}
\newcommand{\tset}{\ensuremath{\mathcal{T}}\xspace}
\newcommand{\rset}{\ensuremath{\mathcal{R}}\xspace}
\newcommand{\kmersof}[1]{\ensuremath{\textrm{\texttt{kmers}}(#1)}\xspace}
\newcommand{\card}[1]{\ensuremath{\left|#1\right|}\xspace}

\newcommand{\sfindex}[2]{\ensuremath{I_{#1}({#2})}\xspace}
\newcommand{\eqClass}[1]{\ensuremath{\left[#1\right]}\xspace}
\newcommand{\arrElem}[2]{\ensuremath{#1\left(#2\right)}\xspace}

\newcommand{\norm}[1]{\ensuremath{\left|\left|#1\right|\right|}}
\newcommand{\kmer}{k-mer\xspace}
\newcommand{\kmers}{k-mers\xspace}
\renewcommand{\bowtie}{Bowtie\xspace}
\newcommand{\rsem}{RSEM\xspace}
\newcommand{\express}{eXpress\xspace}
\newcommand{\cufflinks}{Cufflinks\xspace}
\newcommand{\tophat}{TopHat\xspace}
\newcommand{\sailfish}{Sailfish\xspace}
\newcommand{\suppfigref}[1]{Supplementary Fig.~\ref{#1}\xspace}

\newcommand{\methodseqnref}[1]{Methods, Eqn.~\ref{#1}\xspace}
\newcommand{\methodsalgref}[1]{Methods, Alg.~\ref{#1}\xspace}
\newcounter{countSupNote}
\def\tm{\leavevmode\hbox{$\rm {}^{TM}$}}

\title{\sailfish: Alignment-free Isoform Quantification from RNA-seq Reads using Lightweight Algorithms}

\author[1]{Rob Patro}
\author[2]{Stephen M. Mount}
\author[1]{Carl Kingsford}
\affil[1]{Lane Center for Computational Biology, School of Computer Science \newline
Carnegie Mellon University}
\affil[2]{Department of Cell Biology and Molecular Genetics and Center for Bioinformatics and Computational Biology, University of Maryland}

\date{}

\begin{document}
\maketitle

\begin{abstract}   
RNA-seq has rapidly become the \emph{de facto} technique to measure gene expression.
However, the time required for analysis has not kept up with the pace of data generation.
Here we introduce \sailfish, a novel computational method for quantifying the abundance
of previously annotated RNA isoforms from RNA-seq data.  \sailfish entirely
avoids mapping reads, which is a time-consuming step in all current methods.
\sailfish provides quantification estimates much faster than existing
approaches (typically 20-times faster) without loss of accuracy.
%
\end{abstract}

The ability to generate genomic and transcriptomic data is accelerating beyond
our ability to process it.  The increasingly widespread use and  growing
clinical relevance~(e.g.~\cite{personalizedRNASeq}) of RNA-seq measurements of
transcript abundance will only serve to magnify the divide between our data
acquisition and data analysis capabilities.

The goal of isoform quantification is to determine the relative abundance of
different RNA transcripts given a set of RNA-seq reads. In the analysis of
RNA-seq data, isoform quantification is one of the most computationally time-consuming steps,
and it is commonly the first step in an
analysis of differential expression among multiple
samples~\cite{bmc:diffexp:2013}. There are numerous computational challenges in
estimating transcript-level abundance from RNA-seq data.  Mapping the sequencing 
reads to the genome or transcript sequences can require substantial computational resources.  This often
leads to complicated models that account for read bias and error during 
inference, further adding to the time spent on analysis. Finally, some reads, known as multireads~\cite{rescueStrategy,cufflinks}, 
can map to multiple, sometimes many, different transcripts. The ambiguity resulting 
from these multireads complicates the estimation of relative transcript abundances.

Existing approaches first use read-mapping tools, such as
\bowtie~\cite{bowtie}, to determine potential locations from which the RNA-seq
reads originated.  Given the read alignments, some of the most accurate
transcript quantification tools resolve the relative abundance of transcripts
using expectation-maximization (EM)
procedures~\cite{cufflinks,rsem:2011,express}. In such procedures, reads are
first assigned to transcripts, and these
assignments are then used to estimate transcript abundances.  The abundances
are then used to re-estimate the read assignments, weighting potential matches
in proportion to the currently estimated relative abundances, and these steps
are repeated until convergence.  In practice, both of these steps
can be time consuming.  For example, even when exploiting the
parallel nature of the problem, mapping the reads from a reasonably sized
(\forexample~100M reads) RNA-seq experiment can take hours.

Recent tools, such as eXpress~\cite{express}, aim to reduce the computational
burden of isoform quantification from RNA-seq data by substantially altering the
EM algorithm.  However, even for such
advanced approaches, performing read alignment  and processing the
large number of alignments that result from  ambiguously mapping
reads remains a significant bottleneck and fundamentally  limits the
scalability of approaches that depend on mapping.

\begin{figure}
\begin{center}
\includegraphics[width=0.7\textwidth]{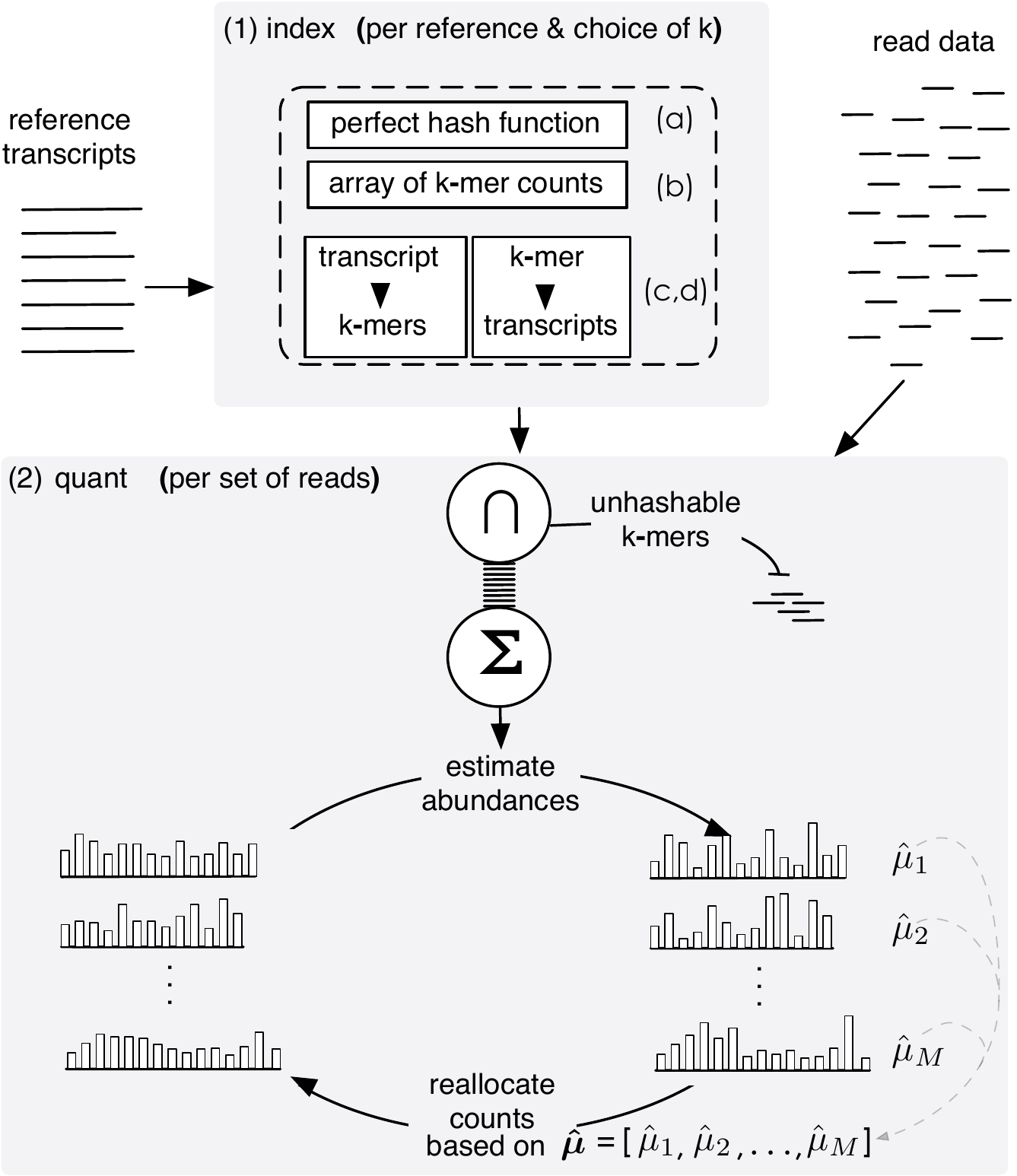}
\caption{\label{fig:pipeline}The \sailfish pipeline consists of an indexing
phase (1) that is invoked via the command \texttt{sailfish index} and a quantification 
phase (2) invoked via the command \texttt{sailfish quant}. The \sailfish index has four
components: (a) a perfect hash function mapping each \kmer in the transcript
set to a unique integer between $0$ and $N$ --- the number of unique \kmers in
the set of transcripts; (b) an array recording the number of times each
\kmer occurs in the reference set; (c) an index mapping each transcript to the
multiset of \kmers that it contains; (d) an index mapping each \kmer to the set
of transcripts in which it appears.  The quantification phase consists of counting
the indexed \kmers in the set of reads and then applying an EM procedure
to determine the maximum-likelihood estimates of relative transcript abundance.}
\end{center}
\end{figure}

\begin{figure}
\begin{subfigure}[b]{\textwidth}
\begin{center}
\begin{tabular}{ll}
\includegraphics[width=0.24\textwidth]{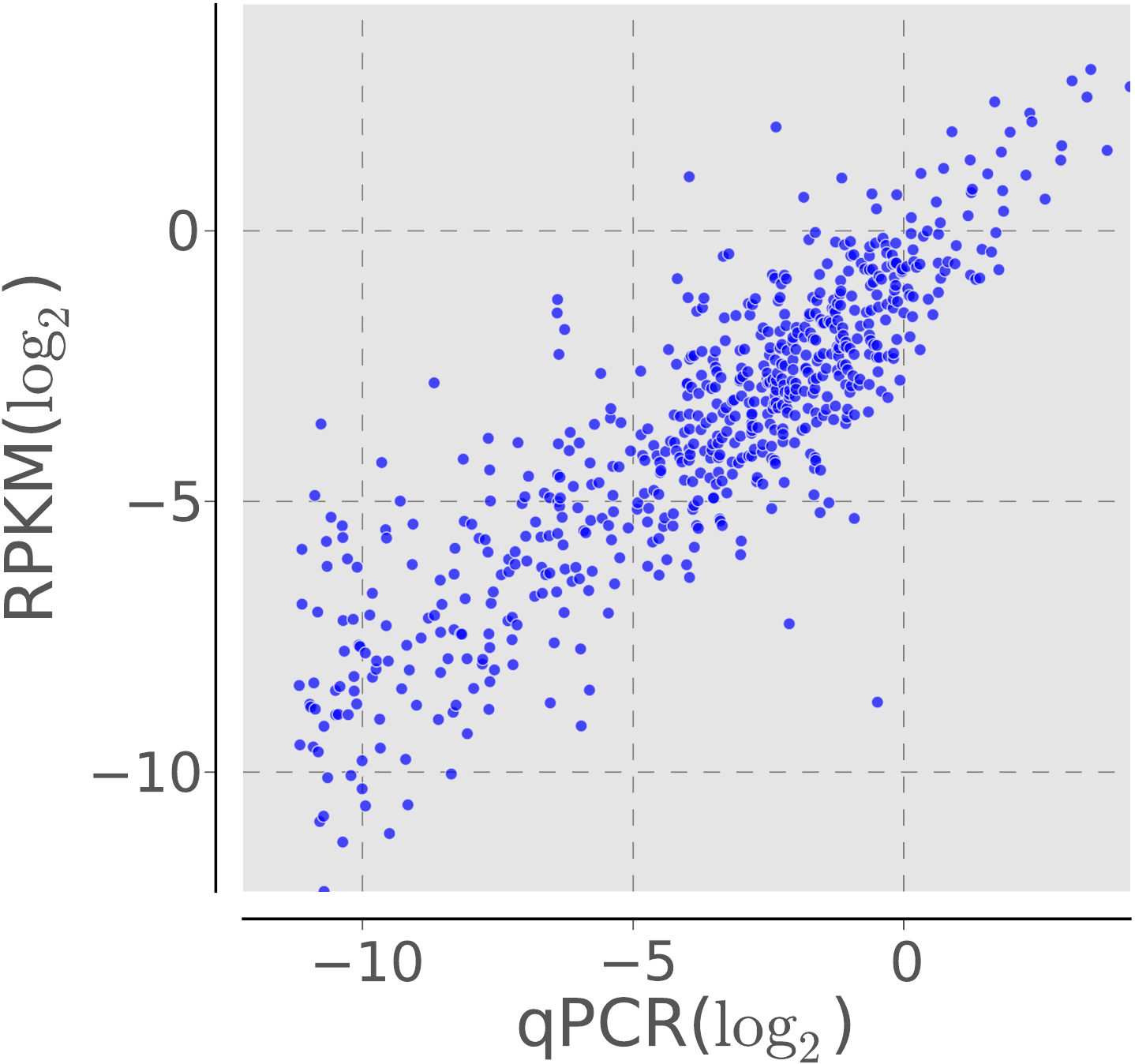} &
\multirow{2}{0.5\textwidth}{\vskip-9em\includegraphics[width=0.53\textwidth]{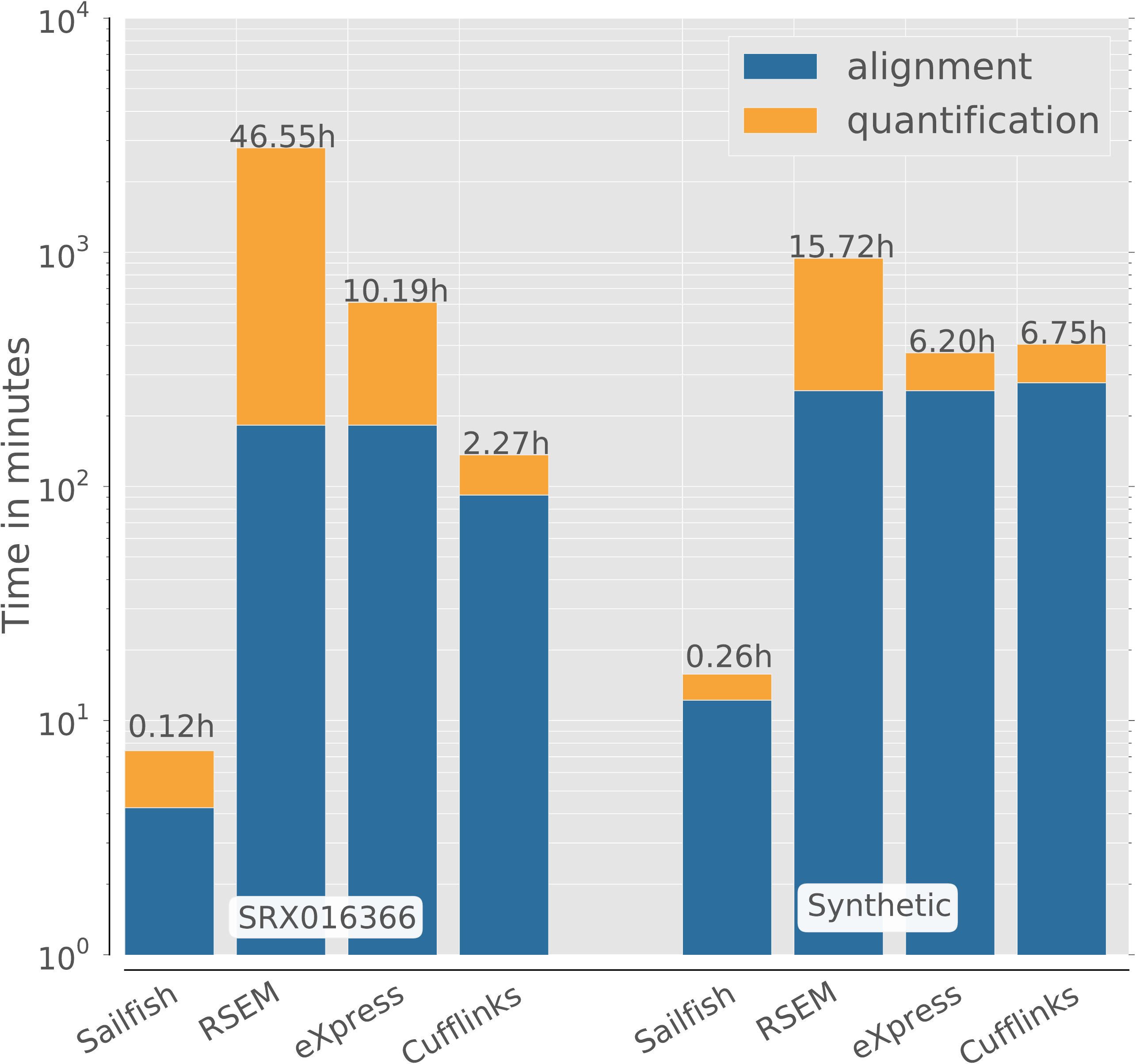}} \\[-9.5em]
\hskip-0.5em \textbf{a} & \multirow{2}{0.5\textwidth}{\vskip-3em\textbf{c}} \\[9.5em]
\includegraphics[width=0.24\textwidth]{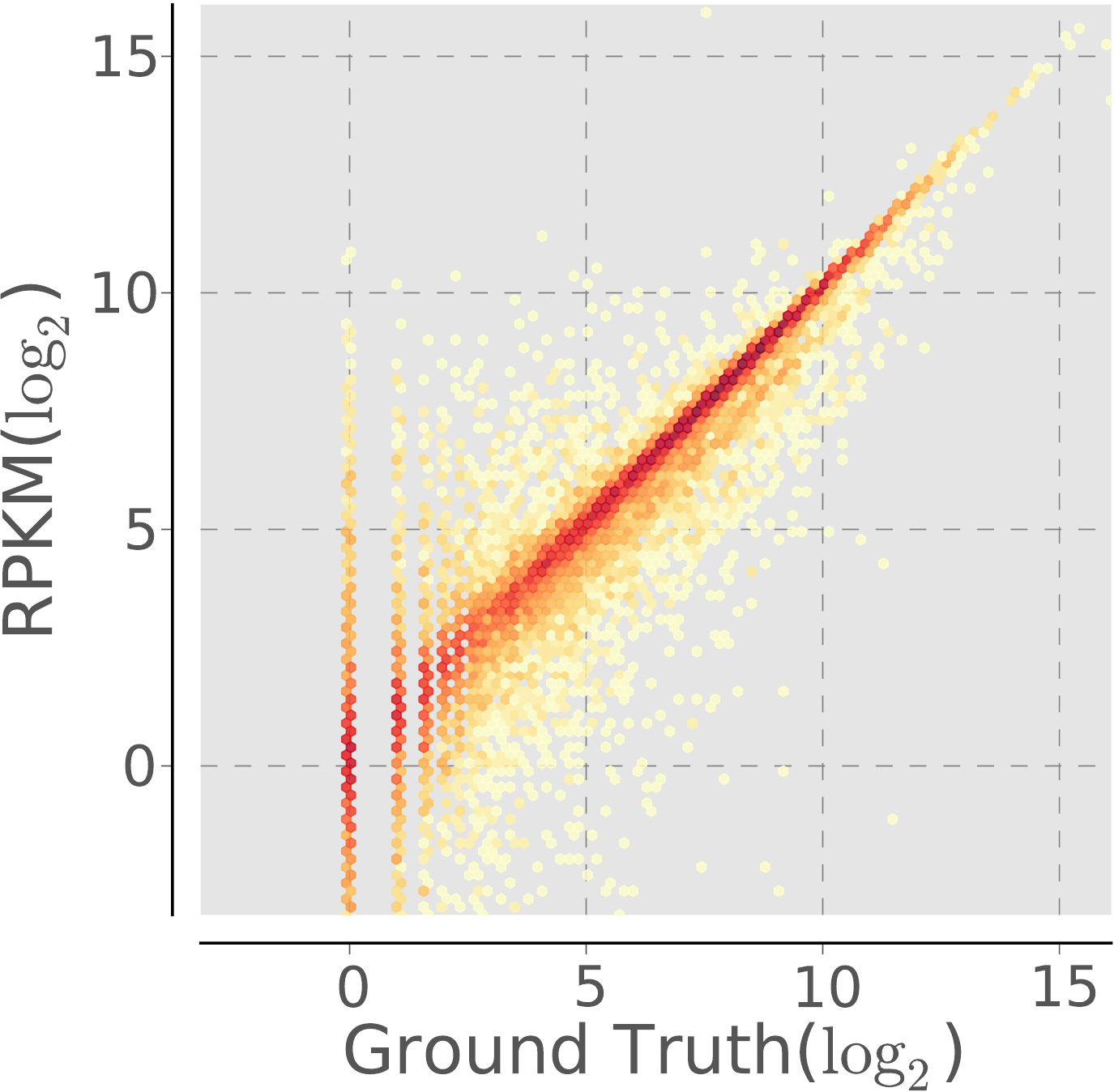} & \\[-9.5em]
\hskip-0.5em \textbf{b} & \\[9.5em]
\end{tabular}
\end{center}
\end{subfigure}

\vskip 0em \hskip 3em \textbf{d}
\vskip -0.5em
\begin{subfigure}[b]{\textwidth}
\begin{center}
{\fontsize{9}{10}\selectfont
\begin{tabular}{ccccccccc}
\toprule
 & \multicolumn{4}{c}{Human Brain Tissue} &
 \multicolumn{4}{c}{Synthetic}\\
 \cmidrule(r){2-5}
 \cmidrule{6-9}
 & \sailfish & \rsem & \express & \cufflinks & \sailfish & \rsem & \express & \cufflinks \\
\midrule
Pearson  & 0.86  & 0.83  & 0.86  & 0.86  & 0.92  & 0.92  & 0.64  & 0.91 \\
Spearman & 0.85  & 0.81  & 0.86  & 0.86  & 0.94  & 0.93  & 0.66  & 0.93 \\
RMSE     & 1.69  & 1.86  & 1.69  & 1.67  & 1.26  & 1.24  & 2.80  & 1.31 \\
medPE    & 31.60 & 36.63 & 32.73 & 30.75 & 4.24  & 5.97  & 26.44 & 6.76 \\
\bottomrule
\end{tabular}}
\end{center}
\phantomsubcaption\label{fig:SyntheticMainResults}
\end{subfigure}
\caption{\label{fig:results} 
(\textbf{a}) The correlation between qPCR estimates of gene abundance (x-axis) and the
estimates of \sailfish. The ground-truth results are taken from the
microarray quality control study (MAQC)~\cite{maqc}. The results shown here
are for the human brain tissue and the RNA-seq based estimates were computed
using the reads from SRA accession \texttt{SRX016366}.  The set of transcripts
used in this experiment were the curated RefSeq~\cite{refseq} genes (accession
prefix NM) from  hg18. (\textbf{b}) The correlation between the
actual number of transcript copies in a simulated dataset (x-axis) and the
abundance estimates of \sailfish.  The transcripts used in this experiment
were  all Ensembl~\cite{ensembl} transcripts from hg19 that were annotated
with a coding feature (CDS).  (\textbf{c}) The total time taken
by each method, \sailfish, \rsem, \express and \cufflinks, to estimate
isoform abundance on each dataset.  The total time taken by a method is the
height of the corresponding bar, and the total is further broken
down into the time taken to perform read-alignment (for \sailfish, we instead
measured the time taken to count the \kmers in the read set) and the time taken
to quantify abundance given the aligned reads (or \kmer counts).  All tools 
were run in multi-threaded mode (where applicable) and were allowed to use
up to $16$ threads.  The table (\textbf{d}) gives the accuracy of each of the methods on both
datasets, as  measured by the Pearson and Spearman correlation coefficients,
root-mean-square error (RMSE) and median percentage error (medPE). 
}
\end{figure}

\sailfish, our software for isoform quantification from RNA-seq data, is
based  on the philosophy of lightweight algorithms, which make frugal
use of data, respect constant factors, and effectively use concurrent hardware
by working with small units of data where possible.
\sailfish avoids mapping reads entirely (Fig.~\ref{fig:pipeline}), resulting in large
savings in time and space. A key technical contribution behind
our approach is the observation that transcript coverage,
which is essential for isoform quantification, can be reliably
and accurately estimated using counts of \kmers occurring in reads. This results in
the ability to obtain accurate quantification estimates more than an order of
magnitude faster than existing approaches, often in minutes instead of hours.
For example, for the data described in Figure~\ref{fig:results}, \sailfish
is between 18 and 29 times faster than the next fastest method while providing
expression estimates of equal accuracy. 

In \sailfish, the fundamental unit of transcript coverage is the \kmer. This is
different from existing approaches, where the fragment or read is the
fundamental unit of coverage. By working with \kmers, we can replace the
computationally intensive step of read mapping with the much faster and
simpler process of \kmer counting. We also avoid any dependence on read
mapping parameters (\forexample~mismatches and gaps) that can have a significant
effect on both the runtime and accuracy of conventional approaches. Yet, our
approach is still able to handle sequencing errors in reads because only the
\kmers that overlap the erroneous bases will be discarded or mis-assigned,
while the rest of the read can be processed as if it were error-free. This
also leads to \sailfish having only a single explicit parameter, the \kmer
length. Longer \kmers may result in less ambiguity, which makes resolving
their origin easier, but may be more affected by errors in the reads.
Conversely, shorter \kmers, though more ambiguous, may be more robust to
errors in the reads (\suppfigref{fig:supinfo:KmerAmbiguity}). Further, we can effectively
exploit modern hardware where multiple cores and reasonably large memories are
common. Many of our data structures can be represented as arrays of
atomic integers (see~\nameref{sec:methods}).  This allows our software to be concurrent and
lock-free where possible, leading to an approach that scales well with the
number of available CPUs (\suppfigref{fig:supinf:CountTimeVsThreads}). Additional
benefits of the Sailfish approach are discussed in Supplementary Note 1.

\sailfish works in two phases: indexing and quantification
(Fig.~\ref{fig:pipeline}).  A \sailfish index is built  from a particular set
of reference transcripts (a \texttt{FASTA} sequence file) and a specific
choice of \kmer length, $k$. The index consists of data structures
that make counting \kmers in a set of reads and resolving their potential
origin in the set of transcripts efficient (see~\nameref{sec:methods}). The
most important data structure in the index is the minimal perfect hash
function~\cite{cmph} that maps each \kmer in the reference transcripts to an
index between $0$ and the number of different \kmers in the transcripts such that
no two \kmers share an index.   This allows
us to quickly index and count any \kmer from the reads that also appears in
the transcripts. We find that pairing the minimum perfect hash function with
an atomically updateable array of \kmer counts allows us to count
\kmers even faster than with  existing advanced lock-free hashes such as that
used in Jellyfish~\cite{jellyfish}. The index also contains a pair of look-up
tables that allow fast access to the indexed \kmers appearing in a specific
transcript as well as the indexed transcripts in which a particular \kmer
appears, both in amortized constant time.  Because the index depends only on
the set of reference transcripts and the choice of \kmer length, it only needs
to be rebuilt when one of these factors changes.

The quantification phase of \sailfish takes as input the index described above
and a set of RNA-seq reads and produces an estimate of the relative abundance
of each transcript in the reference, measured in both Reads Per Kilobase per
Million mapped reads (RPKM) and Transcripts Per Million (TPM);
(see~\nameref{sec:methods} for the definitions of these measures).  First,
\sailfish counts the number of times each indexed \kmer occurs in the set of
reads.  Owing to efficient \kmer indexing by means of the perfect hash
function and the use of a lock-free counting data structure
(\nameref{sec:methods}), this process is efficient and scalable
(\suppfigref{fig:supinf:CountTimeVsThreads}). \sailfish then applies an
expectation-maximization (EM) procedure to determine maximum likelihood
estimates for the relative abundance of each transcript.  Conceptually, this procedure 
is similar to the EM algorithm used by RSEM~\cite{rsem:2011}, except that \kmers
rather than fragments are probabilistically assigned to transcripts, and a two-step
variant of EM is used to speed up convergence. The
estimation procedure first assigns \kmers proportionally to transcripts
(\thatis~if a transcript is the only potential origin for a particular \kmer,
then all observations of that \kmer are attributed to this transcript, whereas
for a \kmer that appears once in each of $n$ different transcripts and occurs $m$ times in the
set of reads, $m/n$ observations are attributed to each potential transcript of origin).  
These initial allocations are then used to estimate the expected
coverage for each transcript (\methodseqnref{eqn:unnormalizedMean}).  In turn, these expected coverage values alter
the assignment probabilities of \kmers to transcripts (\methodseqnref{eqn:allocation}).  
Using these basic EM steps as a building block, we apply a globally-convergent EM
acceleration method, SQUAREM~\cite{squarem}, that substantially increases the convergence rate of the
estimation procedure by modifying the parameter update step and step-length based on
the current solution path and the estimated distance from the fixed 
point (see~\methodsalgref{alg:SQUAREM}).

Additionally, we reduce the number of variables that need to be fit by
the EM procedure by collapsing \kmers into equivalence classes.  Two \kmers
are equivalent from the perspective of the EM algorithm if they occur in the
same set of transcript sequences with the same rate (more details available 
in~\nameref{sec:methods}).  This reduction in the number of active variables
substantially reduces the computational requirements of the EM procedure. For
example, in the set of reference transcripts for which we estimate abundance
using the Microarray Quality Control (MAQC)~\cite{maqc} data 
(Fig.~\ref{fig:results}), there are \num{60504111} \kmers ($k=20$), of which
\num{39393132} appear at least once in the set of reads. However, there are
only \num{151385} distinct equivalence classes of \kmers with non-zero counts.
Thus, our EM procedure needs to optimize the allocations of \num{151385} \kmer
equivalence  classes instead of \num{39393132} individual \kmers, a reduction
by a factor of $\approx260$.

Once the EM procedure converges, the estimated abundances are corrected for
systematic errors due to sequence composition bias and transcript length using
a regression approach similar to Zheng~\andothers~\cite{biascorrect},
though using random forest regression instead of a generalized additive model.
This correction is applied after initial estimates have been produced rather
than at a read mapping or fragment assignment stage, requiring fewer variables
to be fit during bias correction.

To examine the efficiency and accuracy of \sailfish, we compared it to
\rsem~\cite{rsem:2011}, \express~\cite{express} and
\cufflinks~\cite{cufflinks} using both real and synthetic data. Accuracy on
real data was quantified by the agreement between RNA-seq-based
expression estimates computed by each piece of software and qPCR measurements
for the same sample (human brain tissue (HBR) in Fig.~\ref{fig:results} and 
\suppfigref{fig:supinf:qPCRHBRPlots}, and universal human reference tissue (UHR) in \suppfigref{fig:supinf:qPCRUHR}).
These paired RNA-seq and qPCR experiments were performed as part of the
Microarray Quality Control (MAQC) study~\cite{maqc}. qPCR abundance
measurements are given at the resolution of genes rather than
isoforms. Thus, to compare these measurements with the transcript-level
abundance estimates produced by the software, we summed the
estimates for  all isoforms belonging to a gene to obtain an
estimate for that gene. We compare predicted abundances using correlation
coefficients, root-mean-square error (RMSE), and median percentage error
(medPE) (additional details  available in Supplementary Note 2).
Figure~\ref{fig:results} shows that the speed of \sailfish does not sacrifice any
accuracy.

To show that \sailfish is accurate at the isoform level, we generated synthetic data using the Flux
Simulator~\cite{fluxsim}, which allows versatile modeling of various RNA-seq
protocols (see Supplementary Note 3).  Unlike synthetic test
data used in previous work~\cite{rsem:2011,express}, the procedure used
by the Flux Simulator is not based specifically on the generative  model underlying  our
estimation procedure. \sailfish remains accurate at the isoform level (Fig.~\ref{fig:results}).

The memory usage of Sailfish is comparable with that of other tools, using
between 4 and 6 Gb of RAM during isoform quantification for the experiments reported here.


\sailfish applies the idea of lightweight algorithms to the problem of
isoform quantification from RNA-seq  reads and in doing so achieves
a breakthrough in terms of speed.  By eliminating the necessity of read
mapping from the expression estimation pipeline, we not only improve the speed
of the process but also simplify it considerably, eliminating the burden of
choosing all but a single external parameter (the \kmer length) from the user.
As the size and number of RNA-seq experiments grow, we expect \sailfish and its
paradigm to remain efficient for isoform quantification because the memory footprint 
is bounded by the size and complexity of the target transcripts and the only phase that 
grows explicitly in the number of reads --- \kmer counting --- has been designed to 
effectively exploit many CPU cores.

\sailfish is free and open-source software and is available at 
\url{http://www.cs.cmu.edu/~ckingsf/software/sailfish}.

\section*{Methods}
\label{sec:methods}


\paragraph{Indexing.} The first step in the \sailfish pipeline is building an
index from the set of reference transcripts \tset.  Given a \kmer length $k$,
we compute an index $\sfindex{k}{\tset}$ containing four components.  The first
component is a minimum perfect hash function $h$ on the set of \kmers
\kmersof{\tset} contained in \tset.  A minimum perfect hash function is a
bijection between $\kmersof{\tset}$ and the set of integers $\{0,
1, \dots, \card{\kmersof{\tset}}-1\}$.  \sailfish uses the BDZ minimum perfect
hash function~\cite{cmph}. The second component of the index is an
array $C$ containing a count $\arrElem{C}{s_i}$ for every $s_i \in
\kmersof{\tset}$. Finally, the index contains a lookup table $F$, mapping each
transcript to the multiset of \kmers that it contains and a reverse lookup
table $R$ mapping each \kmer to the set of transcripts in which it appears.
The index is a product only of the reference transcripts and the
choice of $k$, and  thus needs only to be recomputed when either of these
changes.

\paragraph{Quantification.} The second step in the \sailfish pipeline is the
quantification of relative transcript abundance; this requires the \sailfish
index $\sfindex{k}{\tset}$ for the reference transcripts $\tset$ as well as a
set of RNA-seq reads $\rset$.  First, we count the number of occurrences of
each $s_i \in \kmersof{\tset} \cap \kmersof{\rset}$.  Since we know exactly
the set of \kmers that need to be counted and already have a perfect hash
function $h$ for this set, we can perform this counting in a particularly
efficient manner. We maintain an array $C_{\mathcal{R}}$ of the appropriate
size \card{\kmersof{\tset}}, where $\arrElem{C_{\mathcal{R}}}{h(s_i)}$
contains the number of times we have thus far observed $s_i$ in \rset.  

Sequencing reads, and hence the \kmers they contain, may originate from
transcripts in either the forward or reverse direction.  To account for both
possibilities, we check both the forward and reverse-complement \kmers from
each read and use a majority-rule heuristic to determine which of the \kmers
to increment in the final array of counts $C_{\mathcal{R}}$.  If the number of
\kmers appearing in $h$ from the forward direction of the read is greater than
the number of reverse-complement \kmers, then we only increment the counts for
\kmers appearing in this read in the forward direction. Otherwise, only counts
for \kmers appearing in the reverse-complement of this read are incremented in
the array of counts.  Ties are broken in favor of the forward directed reads.
By taking advantage of atomic integers and the compare-and-swap (CAS) operation provided
by modern processors, which allows many hardware threads to efficiently update
the value of a memory location without the need for explicit locking, we can
stream through \rset and update the counts of $C_{\mathcal{R}}$ in parallel
while sustaining very little resource contention.

We then apply an expectation-maximization algorithm to obtain estimates of the
relative abundance of each transcript. We define a \kmer equivalence class as
the set of all \kmers that appear in the same set of transcripts with the same
frequency.  In other words, let $\chi(s)$ be a vector that so that entry $t$ of $\chi(s)$ gives how many times
$s$ appears in transcript $t \in \tset$. Then the equivalence class of a \kmer
$s_i$ is given by $\eqClass{s_i} = \{s_j \in \kmersof{\tset} \mid \chi(s_j) =
\chi(s_i)\}$.  When performing the EM procedure, we will allocate counts to
transcripts according to the set of equivalence classes rather than the full
set of transcripts.  We will let $\totCount{\eqClass{s_i}} = \sum_{s_j \in
\eqClass{s_i}} \arrElem{C_{\mathcal{R}}}{h(s_j)}$ denote the total count of
\kmers in \rset that originate from equivalence class $\eqClass{s_i}$.
We say that transcript $t$ contains equivalence class $\eqClass{s}$ if
$\eqClass{s}$ is a subset of the multiset of \kmers of $t$ and denote this
by $\eqClass{s} \subseteq t$.

\paragraph{Estimating abundances via an EM algorithm.} The EM algorithm (Algo.~\ref{alg:EM})
alternates between estimating the fraction of counts of each observed \kmer that
originates from each transcript (E-step) and estimating the relative
abundances of all transcripts given this allocation (M-step).

The E-step of the EM algorithm computes the fraction of each \kmer equivalence
class' total count that is allocated to each transcript.  For equivalence
class \eqClass{s_j} and transcript $t_i$, this value is computed by
\begin{equation}
\alpha(j,i) = \frac{\hat{\mu}_i \totCount{\eqClass{s_j}}}{ \sum_{t \supseteq \eqClass{s_j}} \hat{\mu}_t },
\label{eqn:allocation}
\end{equation}
where $\hat{\mu}_i$ is the currently estimated relative abundance of transcript $i$.
These allocations are then used in the M-step of the algorithm to compute the
relative abundance of each transcript. The relative abundance of transcript
$i$ is estimated by
\begin{equation}
\hat{\mu}_i = \frac{\mu_i}{\sum_{t_j \in \tset} \mu_j},
\label{eqn:normalizedMean}
\end{equation}
where $\mu_i$ is
\begin{equation}
\mu_i = \frac{\sum_{\eqClass{s_j} \subseteq t_i} \alpha(j,i)}{ \hat{l_i} }.
\label{eqn:unnormalizedMean}
\end{equation}
The variable $\hat{l_i}$ denotes the adjusted length of transcript $i$ and is
simply $\hat{l_i} = l_i - k + 1$ where $l_i$ is the length of transcript $i$
in nucleotides.


\begin{algorithm}
\DontPrintSemicolon
\SetKwFunction{EM}{EM}{}  
\SetKw{Def}{function}{}  
\Def\EM{$\bm{\hat{\mu}}$}\;
\Begin{
  \For{$\eqClass{s_j}$}{
    $w = \sum_{t \supseteq \eqClass{s_j}} \hat{\mu}_t$\;
    \For{$t \supseteq \eqClass{s_j}$}{
    $\alpha(j, i) = \hat{\mu}_i \totCount{\eqClass{s_j}} / w$\;
    }
  }

  $y = 0$\;
  \For{$t \in \tset$}{
  $C_t = \sum_{\eqClass{s_j} \subseteq t_i} \alpha(j,i)$\;
  $\mu'_t = C_t /  \hat{l_i}$\;
  $y = y + \mu_t$\;
  }

  \For{$t \in \tset$}{
  $\hat{\mu'}_t = \mu'_t / y$\;
  }

  \Return $\bm{\hat{\mu'}} = \langle \bm{\hat{\mu'}}_0, \bm{\hat{\mu'}}_1, \dots, \bm{\hat{\mu'}}_{\card{\tset}} \rangle$\;
}
\caption{\label{alg:EM}An EM iteration. One iteration of the expectation-maximization
procedure that updates the estimated \kmer allocations $\alpha(\cdot, \cdot)$ and computes new
estimates of relative transcript abundance $\bm{\hat{\mu'}}$ based on the current estimates of 
relative transcript abundance $\bm{\hat{\mu}}$.}
\end{algorithm}

\begin{algorithm}
\DontPrintSemicolon
\SetKwFunction{squarem}{def SQUAREM}{}  
$\bm{\hat{\mu}}_{1} = \mathrm{EM}(\bm{\hat{\mu}}_{0})$\;
$\bm{\hat{\mu}}_{2} = \mathrm{EM}(\bm{\hat{\mu}}_{1})$\;
$r = \bm{\hat{\mu}}_{1} - \bm{\hat{\mu}}_{0}$\;
$v = (\bm{\hat{\mu}}_{2} - \bm{\hat{\mu}}_{1}) - r$\;
$\gamma = -\norm{r} / \norm{v}$\;
modify $\gamma$ by backtracking, if necessary, to ensure global convergence\;
$\bm{\hat{\mu}}_{3} = \max(\bm{0}, \bm{\hat{\mu}}_{0} - 2 \gamma r + \gamma^2 v)$\label{updateline}\;
$\bm{\hat{\mu}}_{0} = \mathrm{EM}(\bm{\hat{\mu}}_{3})$\;
\caption{\label{alg:SQUAREM}A SQUAREM iteration.  Updates the relative abundance
estimates according to an accelerated EM procedure whose update direction and
magnitude are dynamically computed~\cite{squarem}.}
\end{algorithm}
\vspace{1em}

However, rather than perform the standard EM update steps, we perform updates
according to the SQUAREM procedure~\cite{squarem} described in Algo.~\ref{alg:SQUAREM}.
$\bm{\hat{\mu}}_{t} = \langle \hat{\mu}_{0}, \dots, \hat{\mu}_{\card{\tset}}\rangle$ 
is a vector of relative abundance maximum-likelihood 
estimates, and $\textrm{EM}(\cdot)$ is a standard iteration of the
expectation-maximization procedure as outlined in Algo.~\ref{alg:EM}.
For a detailed explanation of the SQUAREM procedure and its proof of convergence, see~\cite{squarem}.  
Intuitively, the SQUAREM procedure builds
an approximation of the Jacobian of $\bm{\hat{\mu}}$ from $3$ successive steps
along the EM solution path, and uses the magnitude of the differences between
these solutions to determine a step size $\gamma$ by which to update the
estimates according to the update rule (line~\ref{updateline}).  The procedure is then
capable of making relatively large updates to the $\bm{\hat{\mu}}$ parameters,
which substantially improves the speed of convergence.  In \sailfish, the
iterative SQUAREM procedure is repeated for a user-specified number of steps
($30$ for all experiments reported in this paper; see~\suppfigref{fig:supinfo:Convergence}).

\paragraph{Bias Correction.}  The bias correction procedure implemented in
\sailfish is based on the model introduced by
Zheng~\andothers~\cite{biascorrect}.  Briefly, it performs a regression
analysis on a set of potential bias factors where the response variables are the estimated
transcript abundances (RPKMs).  \sailfish automatically considers transcript length, GC content 
and dinucleotide frequencies as potential bias factors, as this specific set of features were 
suggested by Zheng~\andothers~\cite{biascorrect}. For each transcript, the prediction of the
regression model represents the contribution of the bias factors to this
transcript's estimated abundance.  Hence, these regression estimates (which may
be positive or negative) are subtracted from the original estimates to obtain
bias-corrected RPKMs.  For further details on this bias correction procedure,
see~\cite{biascorrect}.  The original method used a generalized additive model
for regression; \sailfish implements the approach using random forest
regression to leverage high-performance implementations of this technique. The key
idea here is to do the bias correction after abundance estimation rather than 
earlier in the pipeline.  The bias correction of \sailfish can be disabled with 
the \texttt{--no-bias-correction} command line option. Finally, we note that it
is possible to include other potential features, like normalized coverage plots that can encode 
positional bias, into the bias correction phase.  However, in the current version
of \sailfish, we have not implemented or tested bias correction for these features.

\paragraph{Computing RPKM and TPM.}  \sailfish outputs both Reads Per Kilobase
per Million mapped reads (RPKM) and Transcripts Per Million (TPM) as
quantities predicting the relative abundance of different isoforms.  The RPKM
estimate is the most commonly used, and is ideally $10^9$ times the rate
at which reads are observed at a given position, but the TPM estimate has also
become somewhat common~\cite{rsem:2011,wagner:2012}. Given the relative transcript abundances $\hat{\mu_i}$ 
estimated by the EM procedure described above, the TPM for transcript $i$ is given by
\begin{equation}
\textrm{TPM}_{i} = 10^{6} \hat{\mu}_i.
\label{eqn:TPM}
\end{equation}
Let $C_i = \sum_{\eqClass{s_j} \subseteq t_i} \alpha(j,i)$ be the number of \kmers
mapped to transcript $i$.  Then, the RPKM is given by
\begin{equation}
\textrm{RPKM}_{i} = \frac{\frac{C_i}{\nicefrac{l_i}{10^3}}}{\frac{N}{10^6}} =  \frac{10^9\frac{C_i}{l_i}}{N} \approx  \frac{10^9 \mu_i}{N},
\label{eqn:RPKM}
\end{equation}
where $N = \sum_{\eqClass{s_i}}\totCount{\eqClass{s_i}}$ and the final equality is
approximate only because we replace $l_i$ with $\hat{l_i}$.

\paragraph{Computing Accuracy Metrics.}
Since the RPKM (and TPM) measurements are only \textit{relative} estimates of
isoform abundance, it is essential to put the ground-truth and estimated
relative abundances into the same frame of reference before computing our
validation statistics.  While this centering procedure will not effect
correlation estimates, it is important to perform before computing RMSE and
medPE. Let $X = \{x_i\}_{i=1}^{M}$ denote the ground-truth isoform abundances
and $Y = \{y_i\}_{i=1}^{M}$ denote the estimated abundances.  We transform the
estimated abundances by aligning their centroid with that of the ground-truth
abundances; specifically, we compute the centroid-adjusted abundance estimates
as $Y' = \{y_i - \omega\}_{i=1}^{M}$ where $\omega = \left(\sum_{i=i}^{M} x_i -
\sum_{i=1}^{M} y_i\right) / M$.  It is these centroid adjusted abundance estimates
on which we compute all statistics.

\paragraph{Simulated Data.} The simulated RNA-seq data was generated by the
FluxSimulator~\cite{fluxsim} v1.2  with the parameters listed
in Supplementary Note 3.  This resulted in a dataset of
$150$M, 76 base-pair pared-end reads. \rsem, \express and \cufflinks were
given paired-end alignments since they make special use of this data.
Further, for this dataset, \bowtie~\cite{bowtie} was given the additional flag \texttt{-X990}
when aligning the reads to the transcripts, as $990$ base-pairs was the
maximum observed insert size in the simulated data.
\tophat~\cite{tophat} was provided with the option \texttt{--mate-inner-dist
198}, to adjust the expected mate-pair inner-distance to the simulated
average. The read files were provided directly to \sailfish without any extra
information since the same quantification procedure is used whether single
or paired end reads are provided.

\paragraph{Software Comparisons.} For all comparisons, both \express and \rsem
were provided with the same sets of aligned reads in BAM format.  All reads
were aligned with \bowtie~\cite{bowtie} v0.12.9 using the parameters
\texttt{-aS} and \texttt{-v3}, which allows up to three mismatches per read
and reports all alignments. To prepare an alignment for \cufflinks, \tophat
was run using \bowtie~1 (\texttt{--bowtie1}) and with options \texttt{-N~3} and
\texttt{--read-edit-dist 3} to allow up to three mismatches per read.  For \rsem, 
\express and \cufflinks, the reported times were the sum of the times required for
alignment (via \bowtie for \rsem and \express or via \tophat for \cufflinks) and
the times required for quantification.  The time required for each method
is further decomposed into the times for the alignment and quantification steps
in Fig.~\ref{fig:results}.

\paragraph{Choice of Software Options and Effect on Runtime.} Most expression estimation software,
including \rsem, \express and \cufflinks, provides a myriad of program options
to the user which allow for trade-offs between various desiderata.  For
example, the total time required by \tophat and \cufflinks is lower when
\cufflinks is run without bias correction (\forexample~$1.92$h as opposed to
$2.27$h with bias correction on the \texttt{SRX016366} data). However,
without bias correction, \cufflinks yields slightly lower accuracy (Pearson
$\sigma = 0.82$, Spearman $\rho = 0.81$) than the other methods, while still
taking $16$ times longer to run than \sailfish. Similarly, although aligned
reads can be streamed directly into \express via \bowtie, we empirically
observed lower overall runtimes when aligning reads and quantifying
expressions separately (and in serial), so these times were reported.  Also,
we found that on the synthetic data, the correlations produced by \express
improved to Pearson $\sigma=0.85$, Spearman $\rho=0.87$ and  $\sigma=0.9$,
$\rho=0.92$ when running the EM procedure for 10 or 30 extra rounds
(\texttt{-b~10} and \texttt{-b~30}). However, this also greatly increased the
runtime (just for the estimation step) from 1.9h to 16.71h and 42.12h
respectively. In general, we attempted to run each piece of software with the
options that would be most common in a standard usage scenario.  However,
despite the inherent difficulty of comparing a set of tools  parameterized on
an array of potential options, the core thesis that \sailfish can  provide
accurate expression estimates much faster than any existing tool remains true,
as the fastest performing alternatives, when even when sacrificing accuracy for
speed, were over an order of magnitude slower than \sailfish.

\sailfish version 0.5 was used for all experiments, and all analyses were performed
with a \kmer size of $k=20$.  Bias correction was enabled in all experiments
involving real but not simulated data. The RPKM values reported by \sailfish were 
used as transcript abundance estimates.

\rsem~\cite{rsem:2011} version 1.2.3 was used with default parameters,
apart from being provided the alignment file ($\texttt{--bam}$), for all
experiments, and the RPKM values reported by RSEM were used as abundance
estimates.

\express~\cite{express} version 1.3.1 was used for all
experiments. It was run with default parameters on the MACQ data, and without
bias correction (\texttt{--no-bias-correct}) on the synthetic data. The
abundance estimates were taken as the FPKM values output by eXpress.

\cufflinks~\cite{cufflinks} version 2.1.1 was used for
experiments and was run with bias correction (\texttt{-b}) and multi-read
recovery (\texttt{-u}) on the MACQ data, and with only multi-read recovery
(\texttt{-u}) on the synthetic data.  The FPKM values output by \cufflinks were
used as the transcript abundance estimates.

All experiments were run on a computer with 8~AMD Opteron\tm~6220 processors
(4~cores each) and 256Gb of RAM.  For all experiments, the wall time was
measured using the built-in bash \texttt{time} command.

\paragraph{Implementation of \sailfish.} \sailfish has two basic subcommands,
\texttt{index} and \texttt{quant}. The \texttt{index} command initially builds
a hash of all \kmers in the set of reference transcripts using the
Jellyfish~\cite{jellyfish} software.  This hash is then used to build the
minimum perfect hash, count array, and look-up tables described above. The index
command takes as input a \kmer size via the \texttt{-k} option and a set
of reference transcripts in \texttt{FASTA} format via the \texttt{-t} parameter.
It produces the \sailfish index described above, and it can optionally take
advantage of multiple threads with the target number of threads being provided
via a \texttt{-p} option.

The \texttt{quant} subcommand estimates the relative abundance of transcripts
given a set of reads.  The quant command takes as input a \sailfish index
(computed via the \texttt{index} command described above and provided via the
\texttt{-i} parameter).  Additionally, it requires the set of reads, provided as
a list of \texttt{FASTA} or \texttt{FASTQ} files given by the \texttt{-r}
parameter.  Finally, just as in \texttt{index} command, the \texttt{quant}
command can take advantage of multiple processors, the target number of which is
provided via the \texttt{-p} option.

\sailfish is implemented in C++11 and takes advantage of several C++11 language
and library features.  In particular, \sailfish makes heavy use of built-in
atomic data types.  Parallelization across multiple threads in \sailfish is
accomplished via a combination of the standard library's thread facilities and
the Intel Threading Building Blocks (TBB) library~\cite{IntelTBB}. \sailfish is available as an
open-source program under the GPLv3 license, and has been developed
and tested on Linux and Macintosh OS~X.

\paragraph{Author Contributions} R.P., S.M.M. and C.K. designed the method and algorithms, 
devised the experiments, and wrote the manuscript. R.P. implemented the \sailfish software.

\paragraph{Acknowledgments} This work has been partially funded by National
Science Foundation (CCF-1256087, CCF-1053918, and EF-0849899) and National
Institutes of Health (1R21AI085376 and 1R21HG006913). C.K. received support as
an Alfred P.  Sloan Research Fellow.

\bibliographystyle{abbrv}
\bibliography{SailfishMain}


\newpage

\begin{center}
{\large \singlespacing \sailfish: Alignment-free Isoform Quantification from RNA-seq Reads using Lightweight Algorithms\\}
Rob Patro, Stephen M.Mount and Carl Kingsford
\end{center}

\captionsetup{labelfont=bf}
\setcounter{figure}{0}
\setcounter{page}{1}
\renewcommand{\figurename}{Supplementary Figure}

\newpage

\addtocounter{figure}{1}
\begin{center}
\subsection*{\singlespacing Supplementary Figure \arabic{figure}: Effect of \kmer length on retained data and \kmer ambiguity}
\end{center}
\addtocounter{figure}{-1}
\label{sec:supinf:KmerAmbiguity}
\begin{figure}[H]
\begin{center}
\includegraphics[width=0.75\textwidth]{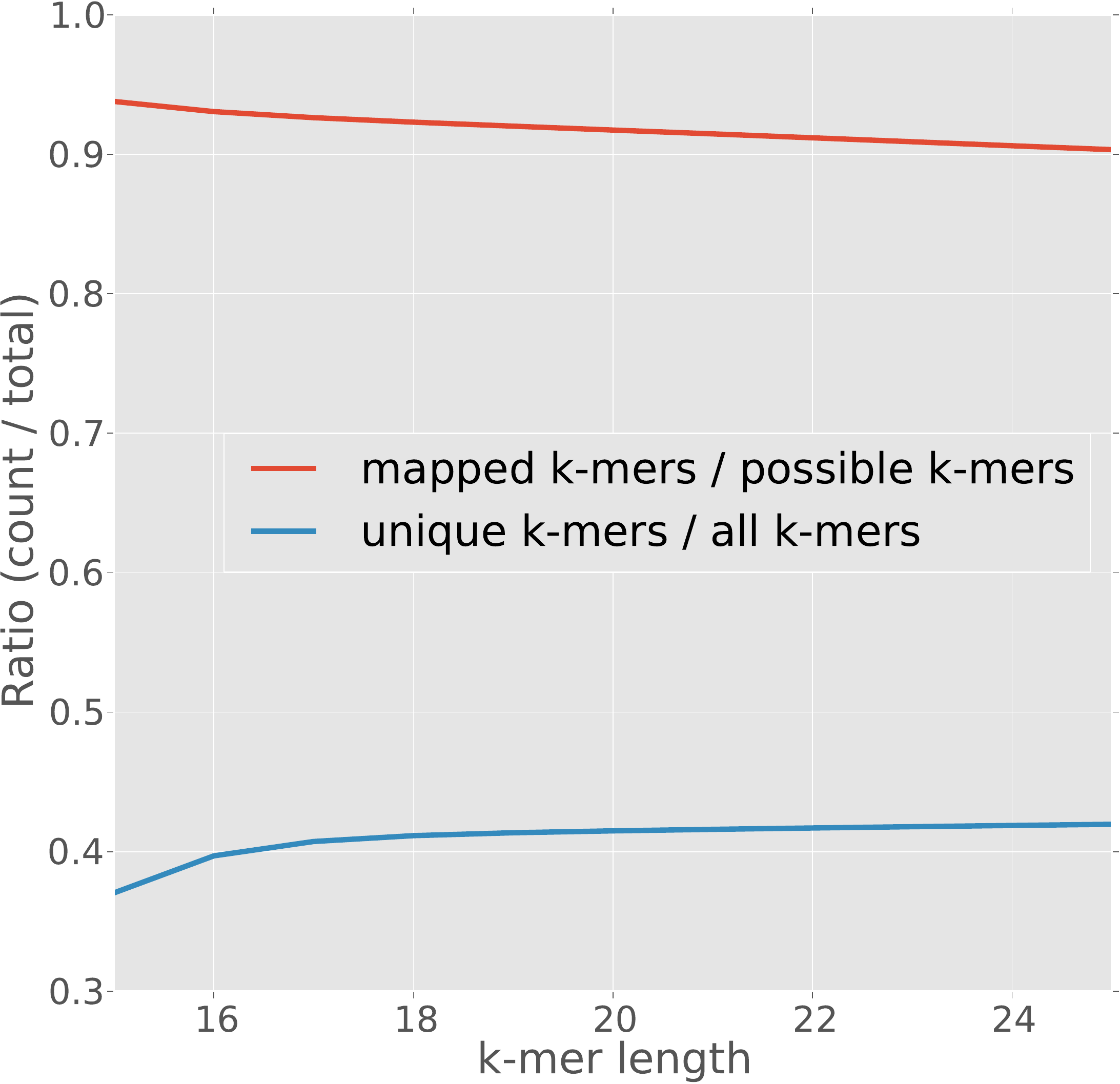}
\caption{\label{fig:supinfo:KmerAmbiguity} As the \kmer length is 
varied in the range $[15,25]$ when processing the synthetic dataset, we observe
that a longer \kmer length results in a slight decrease in data retention (denoted
by the red line which shows the ratio of number of \kmers from the read set that were 
hashable to the total number of \kmers appearing in the set of reads).  Simultaneously,
we observe that the ratio of the number of unique \kmers (\kmers having a unique locus
of origin) in the set of transcripts to the total number of \kmers in the set of transcripts (blue line)
increases as we make $k$ larger.  It seems that, as expected, there is a trade-off 
in the choice of $k$, with a larger $k$ resulting in less robustness to sequencing error but 
a higher fraction unique \kmers and smaller \kmers providing more robustness to errors in
the data but at the cost of increased ambiguity.  However, since the differences are relatively small
over a reasonably large range of $k$, we can expect the inference procedure to 
be fairly robust to this parameter.  We use $k=20$ for all experiments, and this is the default
in \sailfish.  However, we did not attempt to optimize this parameter when performing our experiments.}
\end{center}
\end{figure}

\newpage

\addtocounter{figure}{1}
\begin{center}
\subsection*{Supplementary Figure \arabic{figure}: Speed of counting indexed \kmers}
\end{center}
\addtocounter{figure}{-1}
\label{sec:supinf:CountSpeed}
\begin{figure}[H]
\includegraphics[width=\textwidth]{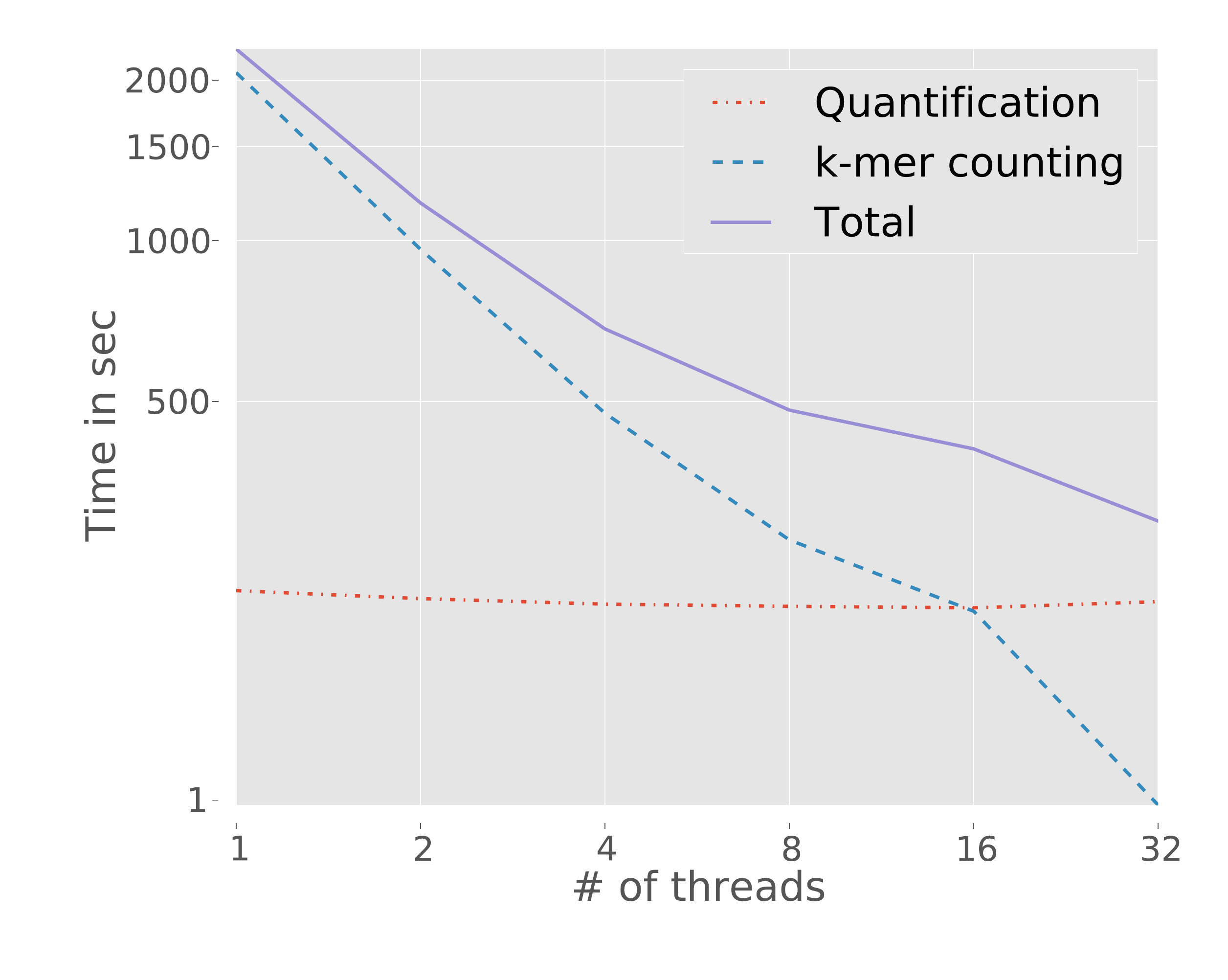}
\caption{\label{fig:supinf:CountTimeVsThreads} The time to count all of the
\kmers and quantify transcript abundance in an $81$M read dataset (\texttt{SRX016366}) as function of the number
of concurrent hashing threads.  Even with only a single thread, the counts for
all \kmers in the dataset can be processed in $34$ minutes and $26$ seconds, while with $32$
processing threads, all \kmers can be counted in only $1$ minute and $28$ seconds.}
\end{figure}

\newpage

\addtocounter{countSupNote}{1}
\begin{center}
\subsection*{Supplementary Note \arabic{countSupNote}: Additional benefits of the \sailfish approach}
\end{center}
\label{sec:supinf:SizeBenefits}
An additional benefit of our lightweight approach is that the size of the
indexing and counting structures required by \sailfish are a small fraction of
the size of the indexing and alignment files required by most other methods.
For example, for the MAQC dataset described in Figure~\ref{fig:results}, the
total size of the indexing and count files required by \sailfish for
quantification was 2.4Gb, compared with much larger indexes and
accompanying alignment files in BAM format used by other approaches (\forexample,
the 15.5Gb index and alignment file produced by \bowtie~\cite{bowtie}).  Unlike alignment files
which grow with the number of reads, the \sailfish index files grow only with
the number of unique \kmers and the complexity of the transcriptome's \kmer composition and
are independent of the number of reads.

\newpage

\addtocounter{figure}{1}
\begin{center}
\subsection*{\singlespacing Supplementary Figure \arabic{figure}: Correlation plots with qPCR on human brain tissue and synthetic data}
\end{center}
\addtocounter{figure}{-1}
\label{sec:supinf:qPCRHBRPlots}

\begin{figure}[H] 
\begin{center}
\begin{tabular}{ccc}
\rsem & \express & \cufflinks \\
\begin{subfigure}[b]{0.33\textwidth}
\includegraphics[width=\textwidth]{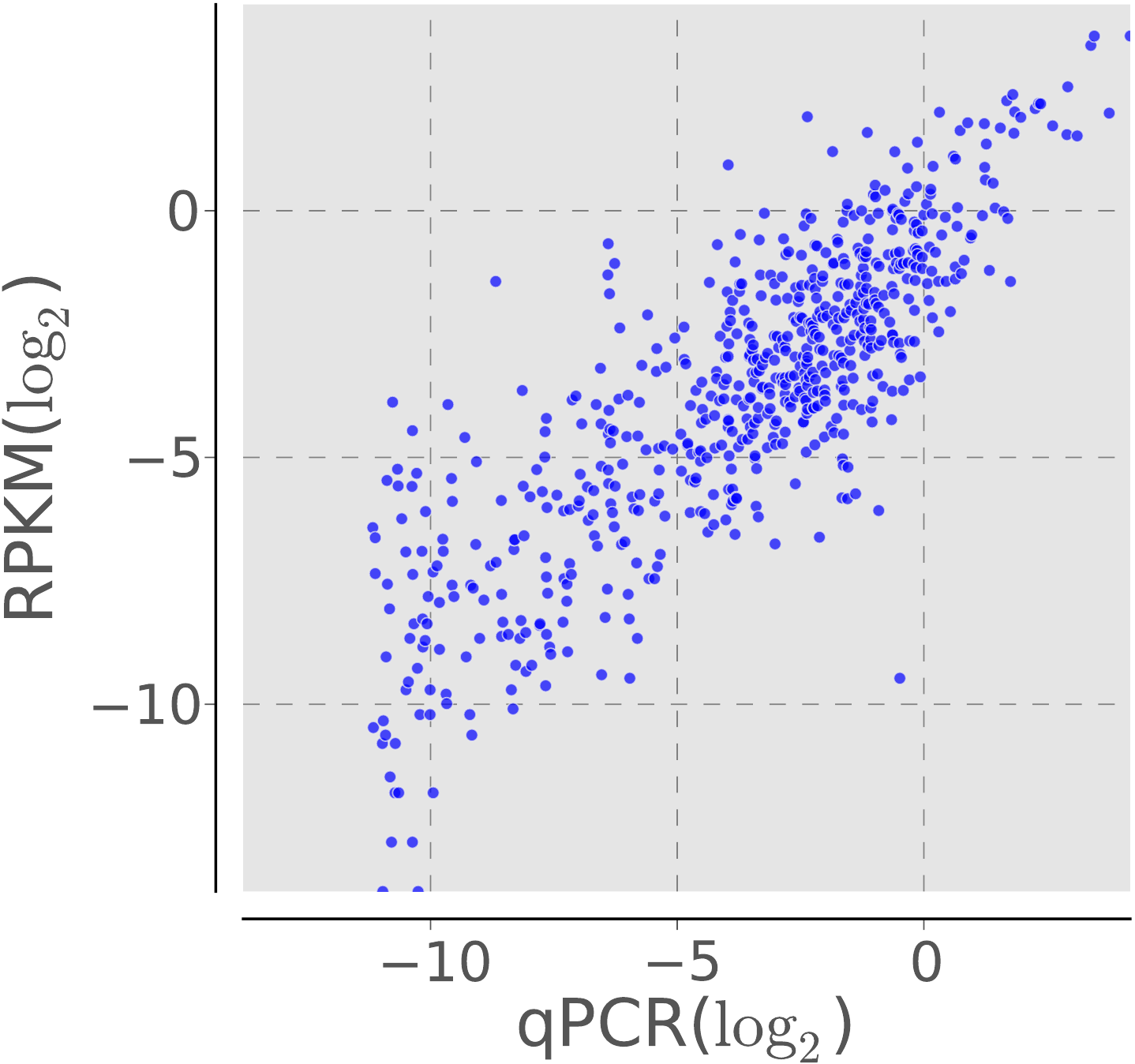}
\end{subfigure} &
\begin{subfigure}[b]{0.3\textwidth}
\includegraphics[width=\textwidth]{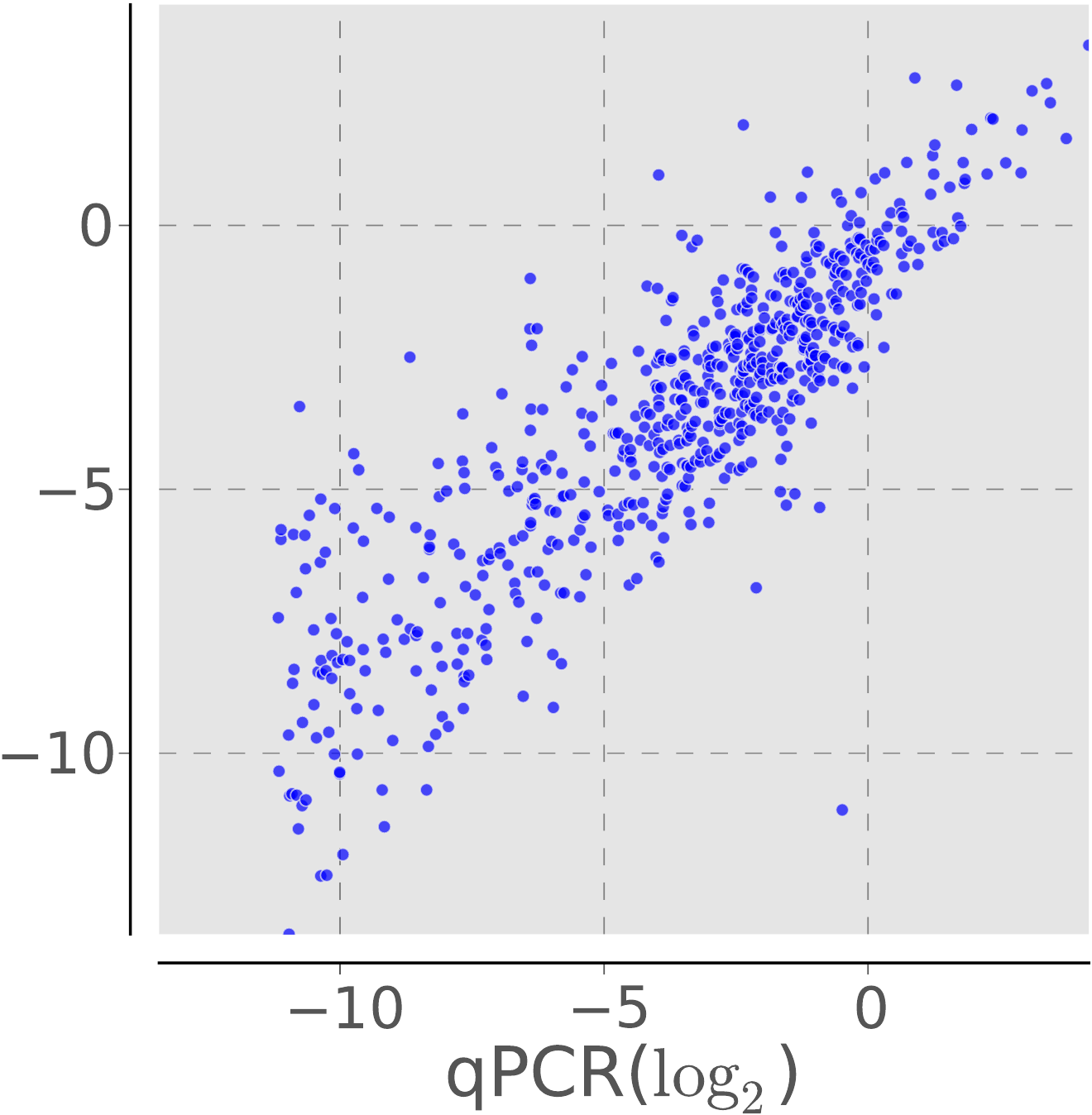}
\end{subfigure} &
\begin{subfigure}[b]{0.3\textwidth}
\includegraphics[width=\textwidth]{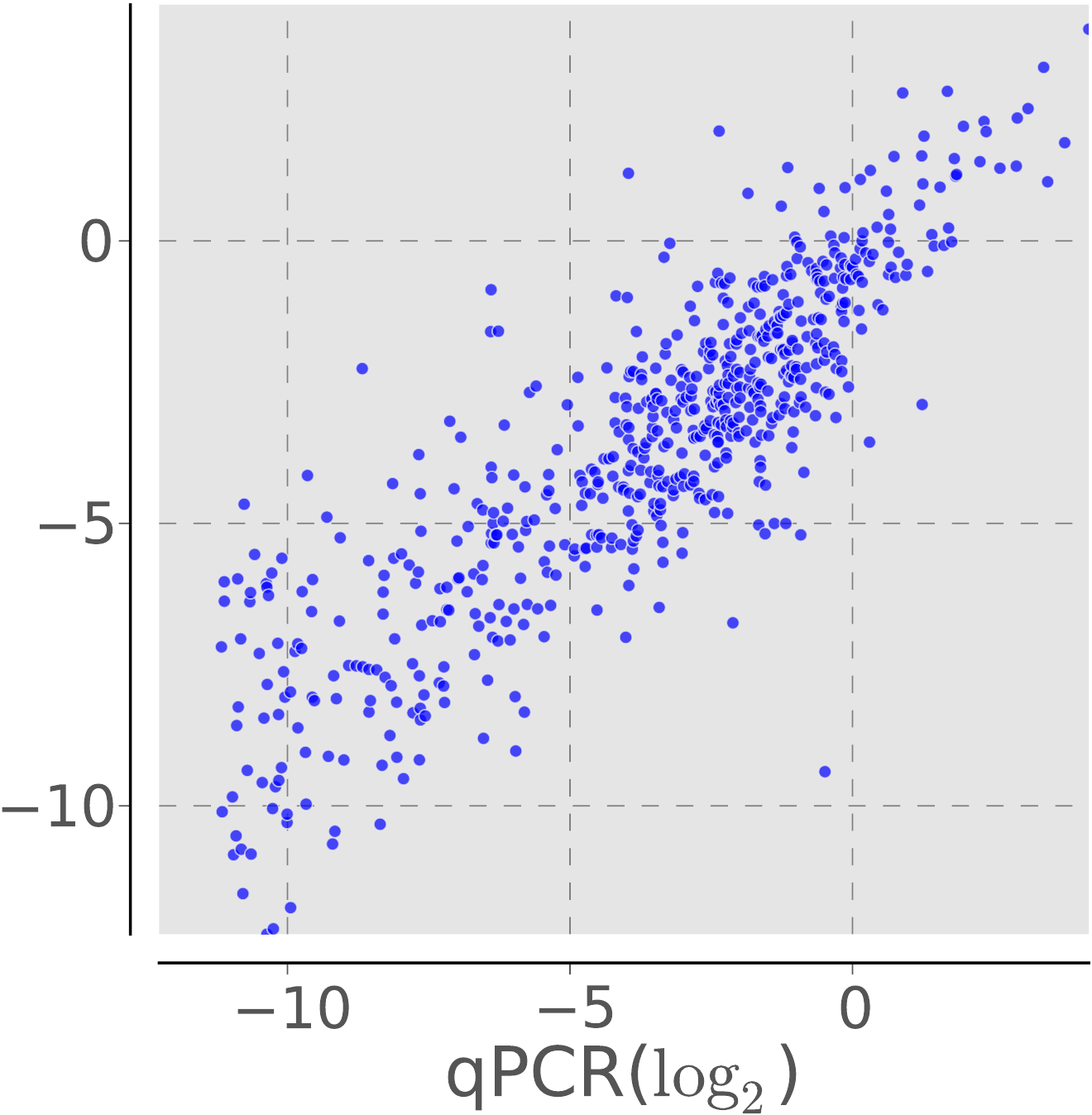}
\end{subfigure} \\

\hskip 0.2em 
\begin{subfigure}[b]{0.33\textwidth}
\includegraphics[width=\textwidth]{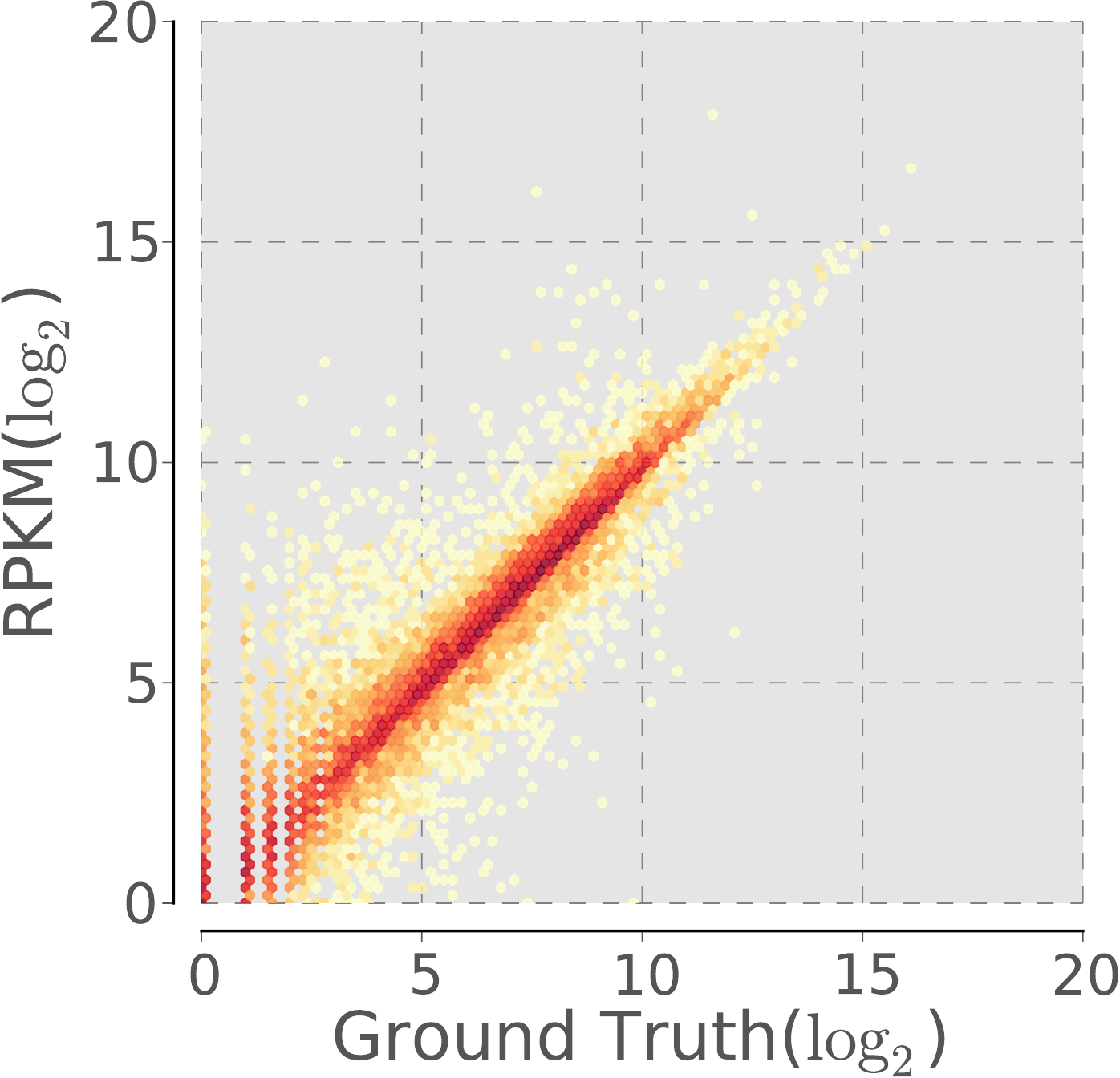}
\end{subfigure} &
\hskip 0.2em 
\begin{subfigure}[b]{0.3\textwidth}
\includegraphics[width=\textwidth]{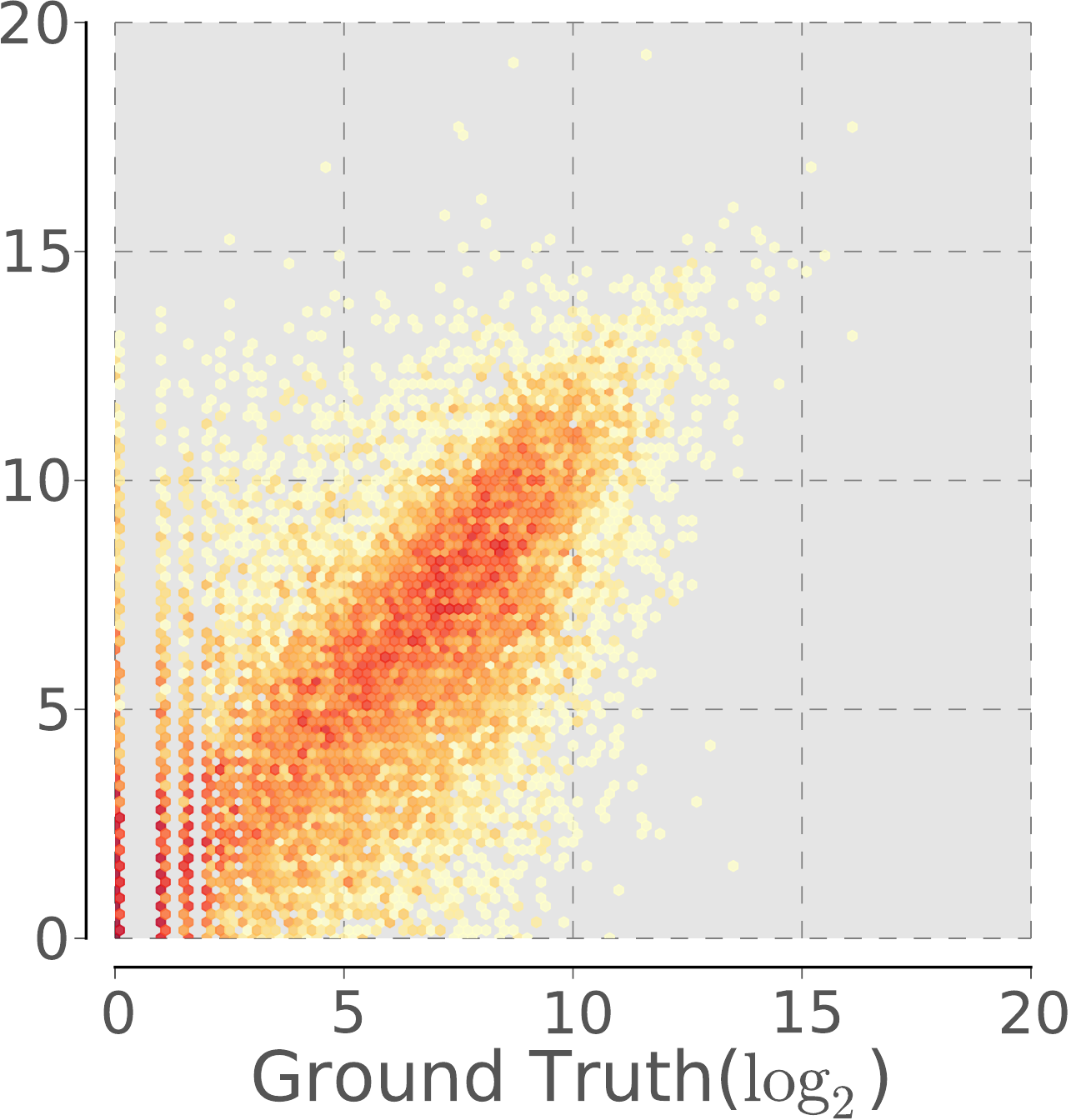}
\end{subfigure} &
\hskip 0.2em 
\begin{subfigure}[b]{0.3\textwidth}
\includegraphics[width=\textwidth]{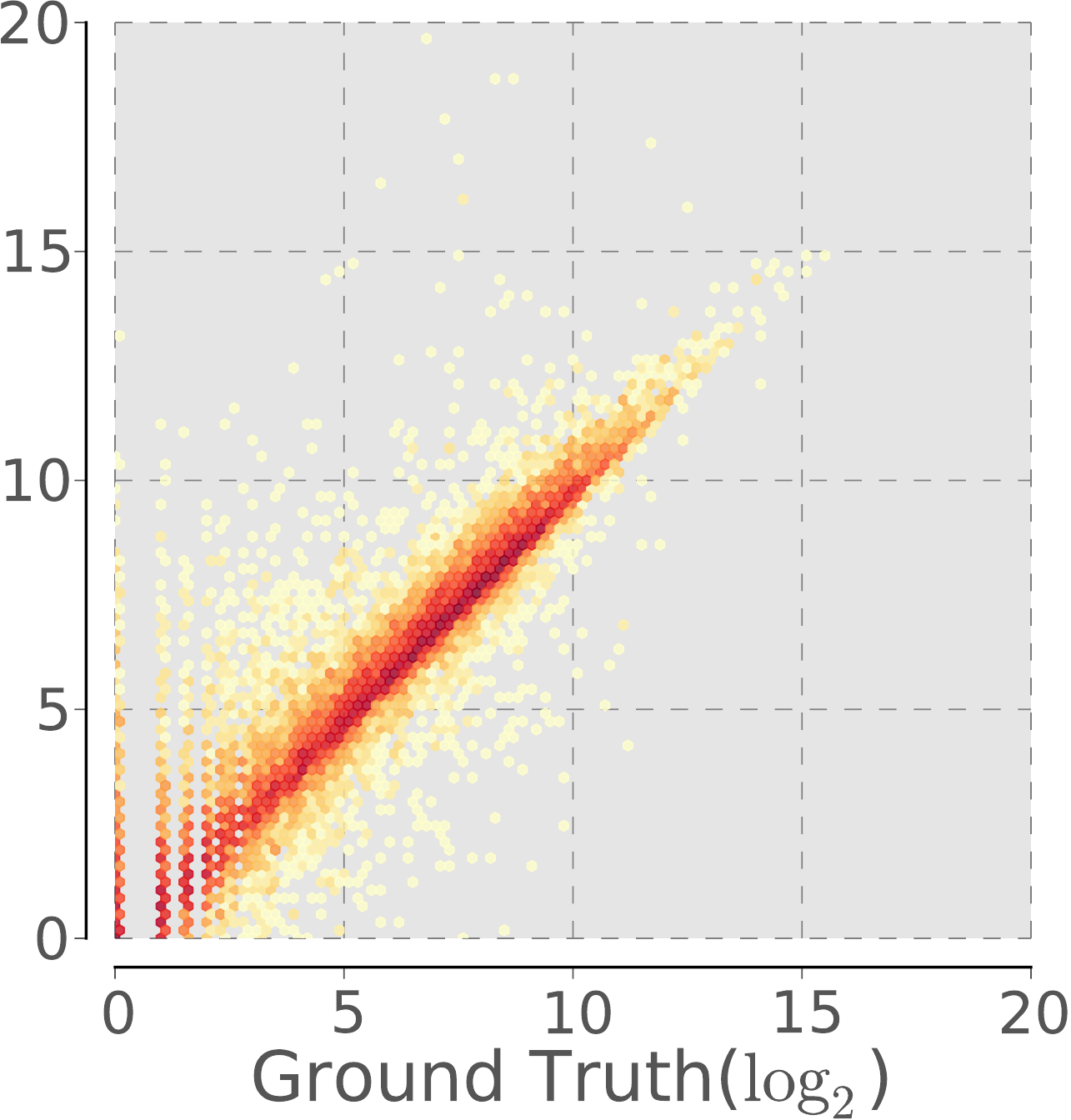}
\end{subfigure}
\end{tabular}

\end{center}
\caption{\label{fig:supinf:qPCRHBRPlots}
Correlation plots of \rsem, \express and \cufflinks for the data
presented in Fig.~\ref{fig:results}.  Each column is labeled with the
method whose output was used to produce that column's plots.  The top
row of plots show the correlation between the computed RPKM and the 
qPCR-based expression estimates for the human brain tissue.  The bottom
row of plots shows the correlation between the computed RPKM and the
true abundance of each transcript on the synthetic dataset.  To generate
the results shown here, \express was run using its default 
streaming expression estimation algorithm.  As reported in~\nameref{sec:methods},
additional batch EM iterations improve \express's accuracy, but come at the cost
of a substantial increase in runtime.}
\end{figure}

\newpage 

\begin{center}
\addtocounter{figure}{1}
\subsection*{\singlespacing Supplementary Figure \arabic{figure}: Correlation with qPCR on universal human reference tissue}
\addtocounter{figure}{-1}
\end{center}
\label{sec:supinf:qPCRUHR}

\begin{figure}[H] 
\begin{center}
\begin{tabular}{cccc}
\sailfish & \rsem & \express & \cufflinks \\
\includegraphics[width=0.22\textwidth]{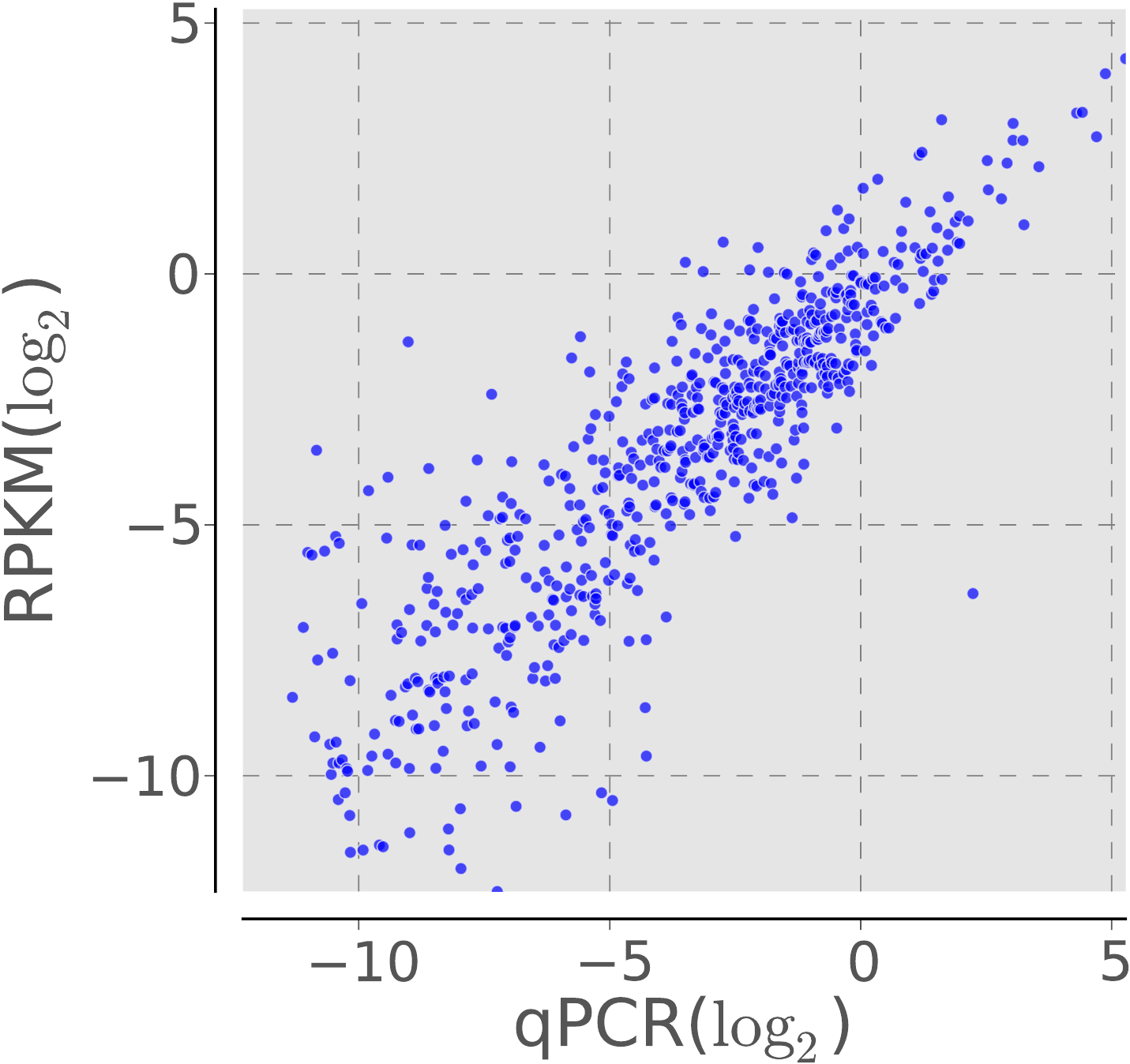} & 
\includegraphics[width=0.20\textwidth]{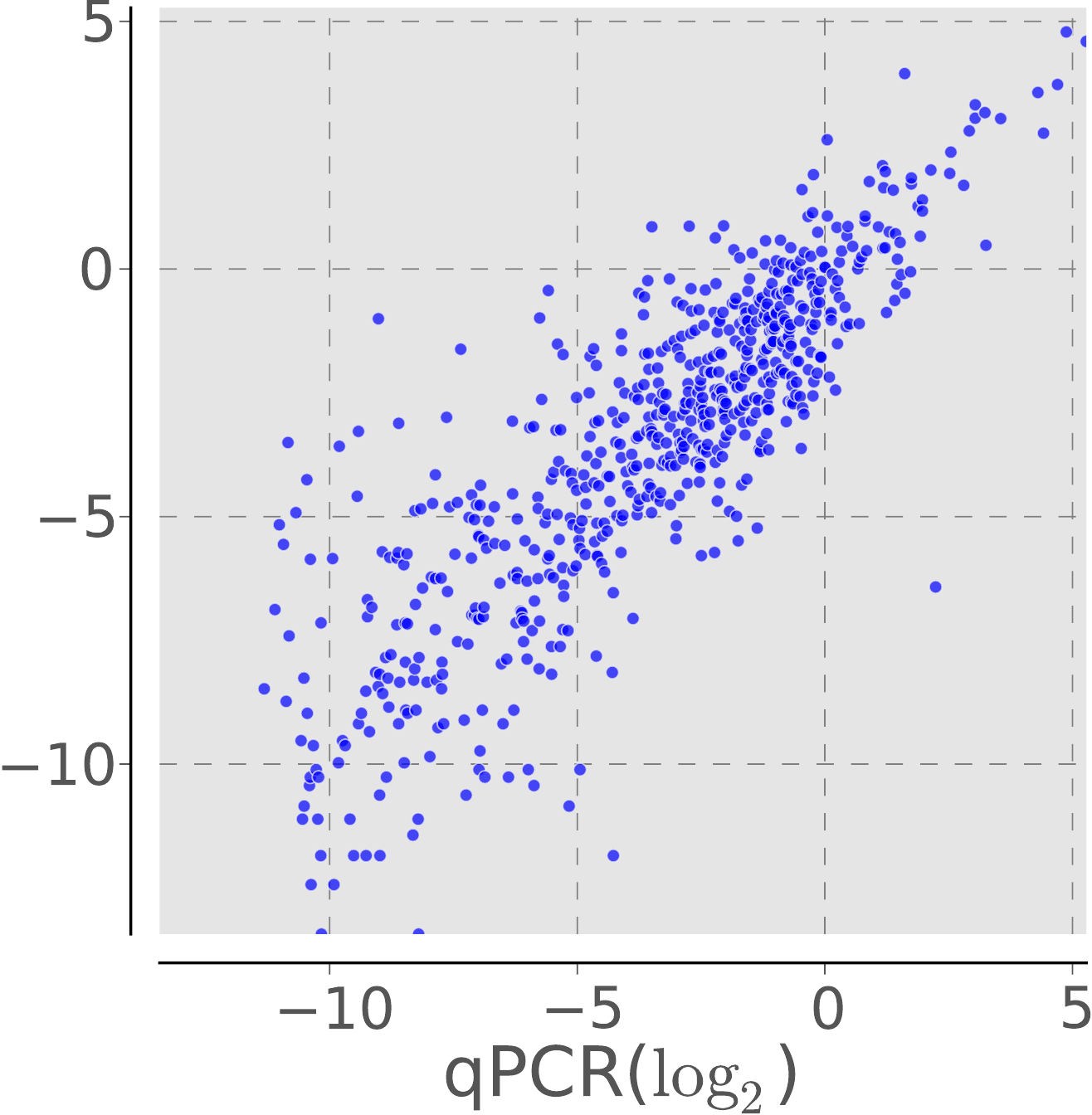} & 
\includegraphics[width=0.20\textwidth]{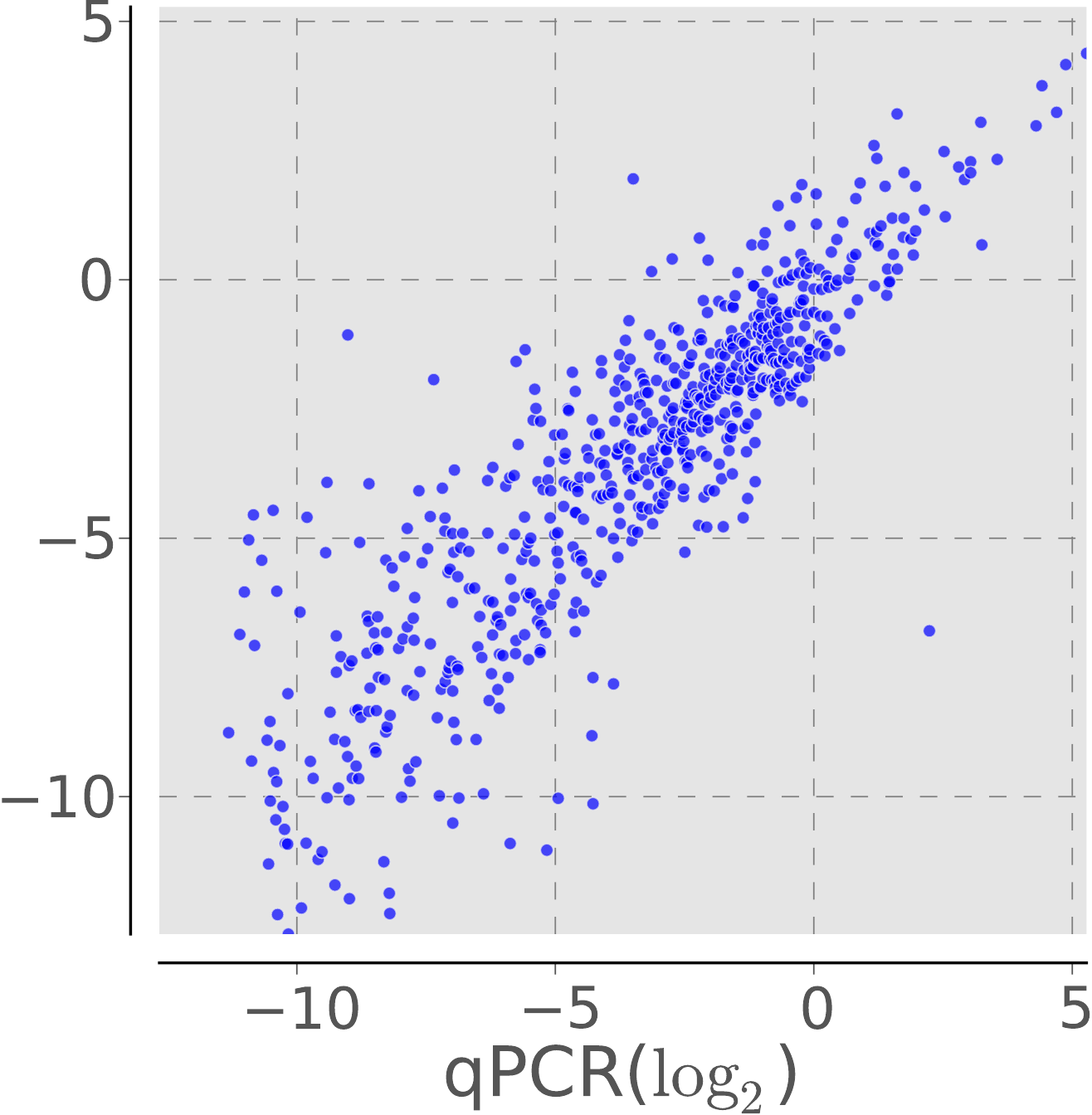} & 
\includegraphics[width=0.20\textwidth]{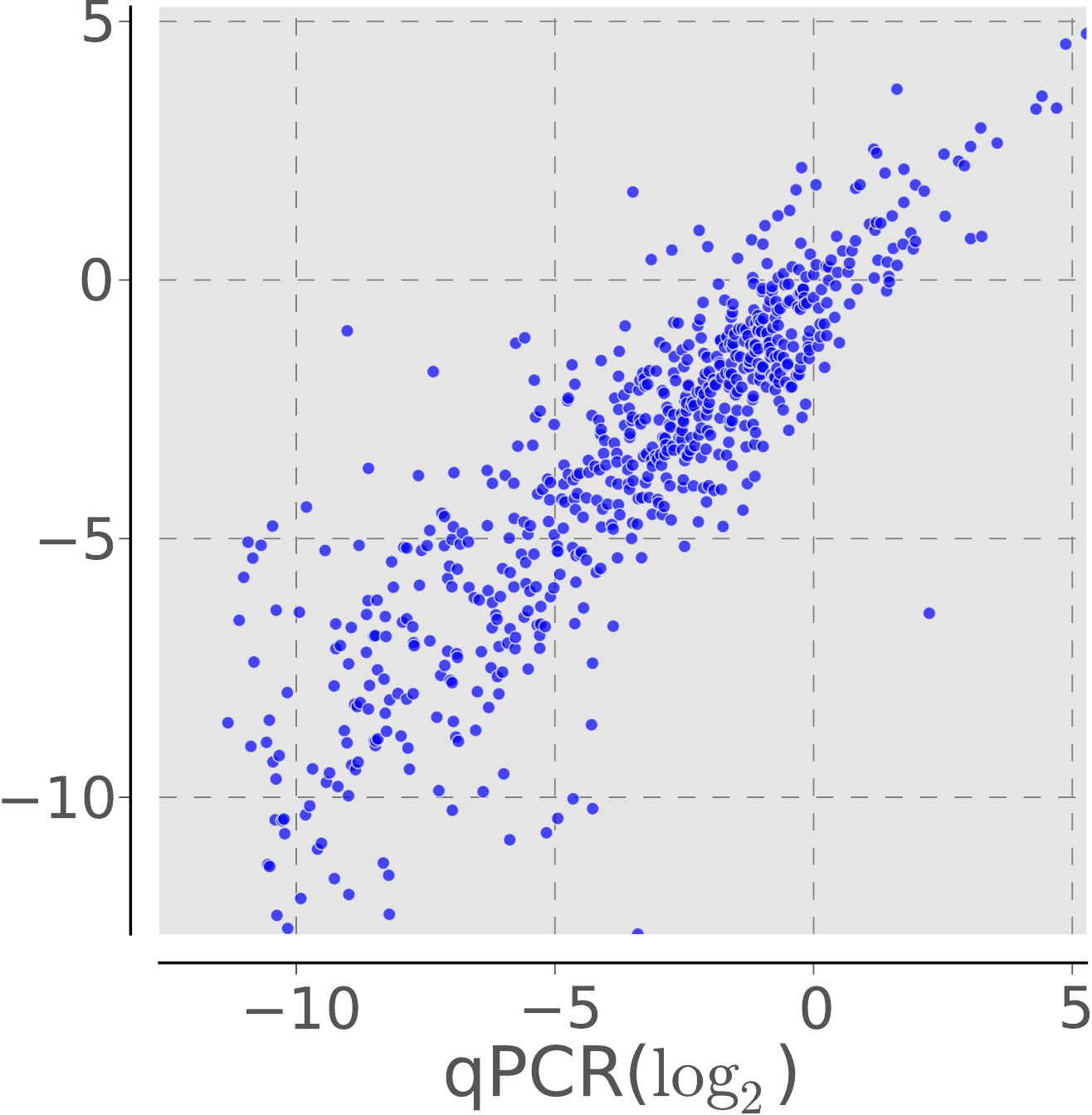} \\
\multicolumn{4}{c}{\includegraphics[width=0.45\textwidth]{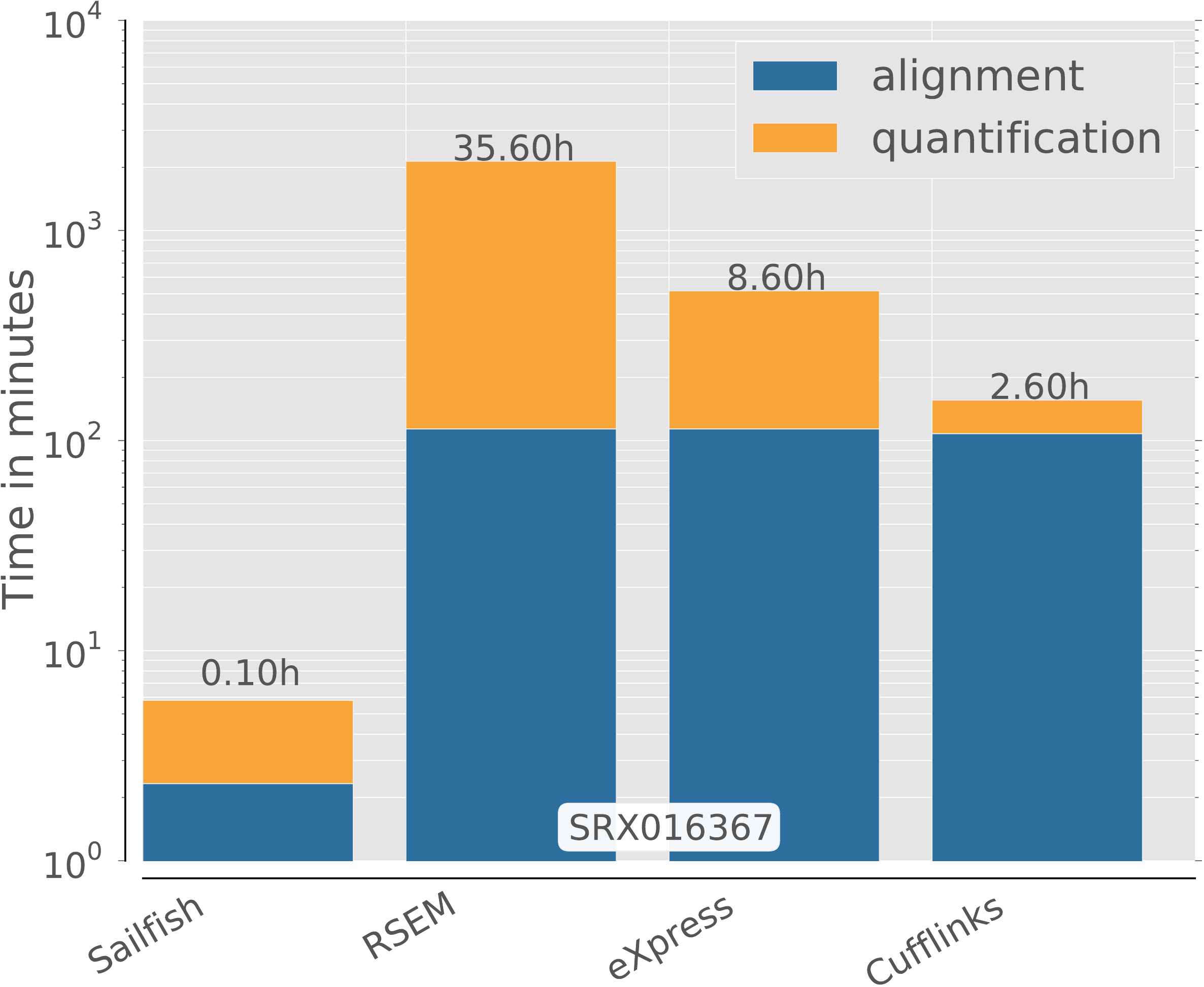}} \\
\end{tabular}

\begin{tabular}{ccccc}
\toprule
 & \sailfish & \rsem & \express & \cufflinks \\
\midrule
Pearson  & 0.87  & 0.85 & 0.87 & 0.87 \\
Spearman & 0.88 & 0.85 & 0.88 & 0.88\\
RMSE  & 1.64 & 1.81 & 1.65 & 1.68\\
medPE & 29.95 & 34.77 & 31.03 & 27.33 \\
\bottomrule
\end{tabular}

\caption{\label{fig:supinf:qPCRUHR}
The accuracy of four methods on a second dataset from
the MACQ~\cite{maqc} study.  The  reads for this experiment were taken from
SRA accession \texttt{SRX016367} ($\approx93$M reads) and are from a mixture
of different tissues (\thatis~the Universal Human Reference or UHR).  The same
set of reference transcripts were used as in Fig.~\ref{fig:results} of the
main text.  The relative accuracy and performance of the methods is similar to
what we observed in the other MACQ dataset, with  \sailfish, \express and
\cufflinks all achieving comparable accuracy (all slightly more accurate than
\rsem).  \sailfish is $\approx26$ times faster then  \cufflinks, the
closest method in terms of speed.}
\end{center}
\end{figure}

\newpage

\begin{center}
\addtocounter{countSupNote}{1}
\subsection*{Supplementary Note \arabic{countSupNote}: Additional details of accuracy analysis}
\end{center}

\label{sec:supinf:AccuracyAnalysis}

We compare predicted abundances using correlation coefficients (Pearson \&
Spearman), root-mean-square error (RMSE), and median percentage error (medPE).
These metrics allow us to gauge the accuracy of methods from different
perspectives . For example, the Pearson correlation coefficient measures how
well trends in the true data are captured by the methods, but, because the
correlation is taken in the log scale, it discounts transcripts with zero 
(or very low) abundance in either sample, while the RMSE includes transcripts 
with true or estimated abundance of zero.  Both \express and \cufflinks produced
a few outlier transcripts, with very low but non-zero estimated abundance,
which significantly degraded the Pearson correlation measure.  We discarded these
outliers by filtering the output of these methods, and setting to zero any estimated
RPKM less than or equal to 0.01, a cutoff chosen because it removed the outliers
but did not seem to discard any other truly expressed transcripts.

\newpage

\begin{center}
\addtocounter{countSupNote}{1}
\subsection*{\singlespacing Supplementary Note \arabic{countSupNote}: Parameters for simulated data}
\end{center}
\label{sec:supinf:SyntheticParams}
The simulated RNA-seq data was generated by the FluxSimulator~\cite{fluxsim} v1.2 with
the following parameters.

{\renewcommand\baselinestretch{1}
\begin{verbatim}
### Expression ###
NB_MOLECULES    5000000
REF_FILE_NAME   hg19_annotations.gtf
GEN_DIR         hg19/chrs
LOAD_NONCODING  NO
TSS_MEAN        50
POLYA_SCALE     NaN
POLYA_SHAPE     NaN
### Fragmentation ###
FRAG_SUBSTRATE  RNA
FRAG_METHOD     UR
FRAG_UR_ETA     350
FRAG_UR_D0      1
### Reverse Transcription ###
RTRANSCRIPTION  YES
RT_PRIMER       RH
RT_LOSSLESS     YES
RT_MIN  500
RT_MAX  5500
### Amplification ###
GC_MEAN NaN
PCR_PROBABILITY 0.05
FILTERING       NO
### Sequencing ###
READ_NUMBER     150000000
READ_LENGTH     76
PAIRED_END      YES
ERR_FILE        76
FASTA           YES
UNIQUE_IDS      YES
\end{verbatim}
}

\newpage

\begin{center}
\addtocounter{figure}{1}
\subsection*{Supplementary Figure \arabic{figure}: Convergence of relative abundance estimates}
\addtocounter{figure}{-1}
\end{center}
\label{sec:supinf:Convergence}
\begin{figure}[H]
\begin{center}
\includegraphics[width=\textwidth]{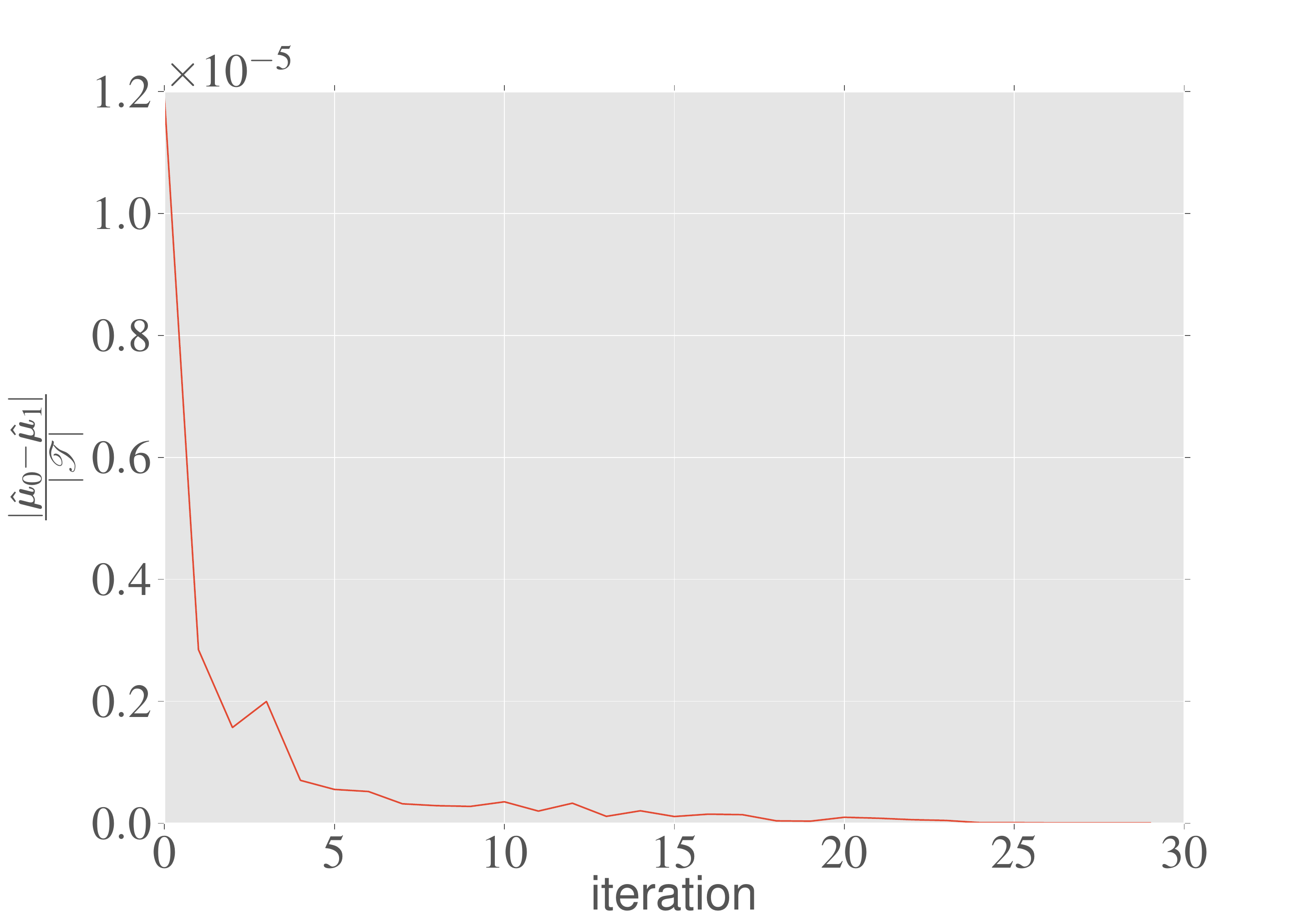}
\caption{\label{fig:supinfo:Convergence} The average difference between the relative abundance
as estimated by two successive applications of the EM step (Algo.~\ref{alg:SQUAREM} lines 1--2)
versus iterations of the SQUAREM algorithm (in the Universal Human Reference tissue experiment).  
We can see that the residual drops off quickly, and appears to have converged before 30 iterations 
of the SQUAREM procedure have been performed.}
\end{center}
\end{figure}

\end{document}